\documentclass[aps,prd,twocolumn,unsortedaddress,showpacs,nofootinbib]{revtex4-1}

\usepackage{amsmath,amssymb,graphicx}
\usepackage{color}
\usepackage{pslatex}
\usepackage{pdfpages}
\usepackage{graphicx}
\usepackage{psfrag}

\usepackage{graphicx}
\usepackage{hyperref}

\usepackage[normalem]{ulem}

\begin{document}

\title{Ruling out thermal dark matter with a black hole induced spiky profile in the M87 galaxy}

\author{Thomas Lacroix}
\affiliation{Institut d'Astrophysique de Paris, UMR 7095, CNRS, UPMC Universit\'{e} Paris 6, Sorbonne Universit\'{e}s, 98 bis boulevard Arago, 75014 Paris, France}
\email{lacroix@iap.fr}
\author{C\'{e}line B\oe hm}
\affiliation{Institute for Particle Physics Phenomenology, Durham University, Durham, DH1 3LE, United Kingdom}
\affiliation{LAPTH, Universit\'{e} de Savoie, CNRS, BP 110, 74941 Annecy-Le-Vieux, France}
\email{c.m.boehm@durham.ac.uk}
\author{Joseph Silk}
\affiliation{Institut d'Astrophysique de Paris, UMR 7095, CNRS, UPMC Universit\'{e} Paris 6, Sorbonne Universit\'{e}s, 98 bis boulevard Arago, 75014 Paris, France;}
\affiliation{The Johns Hopkins University, Department of Physics and Astronomy,
3400 N. Charles Street, Baltimore, Maryland 21218, USA}
\affiliation{Beecroft Institute of Particle Astrophysics and Cosmology, Department of Physics,
University of Oxford, Denys Wilkinson Building, 1 Keble Road, Oxford OX1 3RH, United Kingdom}
\email{silk@iap.fr}

\date{\today}

\begin{abstract}
Using the spectral energy distribution of M87, a nearby radio galaxy in the Virgo cluster, and assuming a supermassive black hole induced spike in the dark matter halo profile,  we exclude  any dark matter candidate with a velocity-independent (s-wave) annihilation cross-section of the order of $\left\langle \sigma v \right\rangle \sim 10^{-26} \ \rm{cm^3\ s^{-1}}$ and a mass up to $O(100)$ TeV.  These limits supersede all previous constraints on thermal, s-wave, annihilating dark matter candidates by orders of magnitude, and  rule out the entire canonical mass range. We remark in addition that, under the assumption of a spike, dark matter particles with  a mass of a few TeV and an annihilation cross-section of $\sim 10^{-27} \ \rm{cm^3\ s^{-1}}$ could explain the TeV $\gamma$-ray emission observed in M87. A central dark matter spike is plausibly present around the supermassive black hole at the center of M87, for various, although not all, formation scenarios, and would have profound implications for our understanding of the dark matter microphysics.

\end{abstract}


\maketitle

\section{Introduction}

Finding signatures of dark matter annihilations in the sky has always been a  priority for the dark matter community. Not only would this validate the particle nature of dark matter (DM) but it would also give some insights about dark matter properties. Evidence for an excess of cosmic-rays in a DM-rich environment (in particular our Galactic center) could provide indirect evidence for DM annihilations or decay. Evidence for a total velocity-independent annihilation cross-section of about $\left\langle \sigma v \right\rangle \simeq 3 \times 10^{-26} \ \rm{cm^3\ s^{-1}}$ would in addition support the hypothesis that DM was once in thermal equilibrium with standard model particles. 

In the absence of annihilation signatures in DM halos, stringent limits are being placed on the DM self-annihilation cross-section as a function of the DM mass, $m_{\rm{DM}}$. The most severe limits originate from e.g. the diffuse $\gamma$-ray background in the Milky Way and its companion galaxies (dwarf spheroidals), as well as from distortions of the cosmic microwave background (CMB) and both high energy positrons and antiprotons. Altogether these measurements already rule out the simplest\footnote{One exception to this conclusion being scenarios with coannihilations, see e.g.~\cite{Griest:1990kh}.} thermal, velocity independent, dark matter scenarios with a mass ranging from a few MeV \cite{Boehm2004,Lopez-Honorez2013} to $\sim$ 100 GeV (see e.g.~\cite{FLATC2015,PlanckCollaboration2015,Spanish_group,Bergstroem2013,Ibarra2014,Bringmann_constraints,Giesen2015}), but here we show that one can go a step further. We use the spectral diffuse emission of M87, a nearby radio galaxy in the Virgo cluster located about 16 Mpc from us, to exclude heavier and more weakly interacting DM particles.

We observe that the presence of a supermassive black hole (BH) in the core of M87 may increase the DM energy distribution so much toward the galactic center that the predicted flux expected from thermal DM particles would exceed observations. In the standard picture, the DM energy density follows  a power law, $\rho \propto r^{-\gamma}$ in the inner region, with $r$ the distance from the galactic center and $\gamma \sim 1$ the slope for a Navarro-Frenk-White (NFW) profile~\cite{NFW}. However, very close to the black hole, the DM profile may rise very steeply. Assuming that the BH grew adiabatically,  expectations are that $\gamma$ should instead lie between 2.25 and 2.5, with the typical value $ \gamma \equiv \gamma_{\mathrm{sp}} = 7/3$, according to Ref.~\citep{spikeGS}, in the very inner region. Such an enhancement of the DM energy density is referred to as a spike and is expected to enhance the brightness of the electromagnetic flux originating  from annihilating DM particles by a factor $(\rho_{\mathrm{sp}}/\rho_{\mathrm{NFW}})^2 \sim r^{\, 2 \, (1-\gamma_{\mathrm{sp}})}$, at least close to the DM mass threshold. 

The existence of such a DM spike is however debatable. If the growth of the BH was not adiabatic (as expected if the BH seed were brought in by a merger), then the inner DM energy density profile would behave instead as $\rho \propto r^{-4/3}$~\citep{Ullio2001,Gnedin2004}. Besides, if the DM halo itself underwent a merger, or if the BH did not grow exactly at the center of the DM halo, the inner DM halo profile would follow $\rho \propto r^{-1/2}$, thus considerably reducing the electromagnetic flux expected from DM particles. Finally, even if a spike could form with $\gamma_{\rm{sp}} \sim 7/3$, the process of dynamical relaxation by DM scattering off stars could smooth down the spike and lead to a DM halo profile of the form $\rho \propto r^{-3/2}$, which would correspond to a Moore profile \cite{Moore1999}.

The latter argument may not be relevant for M87, though. Indeed dynamical heating by stars is inefficient when the dynamical relaxation time $t_{\mathrm{r}}$ in the core is larger than the Hubble time ($\sim10^{10}\ \rm yr$). This effect varies significantly from one galaxy to another, depending on the dynamical properties of the stellar core \citep{Vasiliev2008}. M87 is dynamically young: its relaxation time is estimated to be $t_{\mathrm{r}} \sim 10^{5}$ Gyr (instead of $\sim 2.5$ Gyr for the Milky Way) due to the strong dependence on the velocity dispersion of the stars and DM. \footnote{The relaxation time goes as $t_{\mathrm{r}} \propto \sigma^{3}$ \citep{Vasiliev2008,Binney1987}, with $\sigma$ proportional to $M_{\mathrm{BH}}^{1/2}$, where $M_{\mathrm{BH}}$ is the mass of the BH.} Hence, a spike formed at early times is much more likely to have survived galaxy dynamics up to the present epoch in M87 than in the Milky Way. 

In what follows, we assume that a spike has formed in M87 and, given the above argument, also survived the scattering off stars. We will study its impact on the electromagnetic signatures expected from DM annihilations and derive stringent limits on the DM properties. The paper is structured as follows. First we review the calculations of the electromagnetic flux from annihilating DM in Sec.~\ref{DM_spike_only}. Then we derive in Sec.~\ref{DM_upperlimits} the upper limits on the annihilation cross-section as a function of  the DM mass in the presence or absence of a spike. In Sec.~\ref{spike+jet}, we find the corresponding upper limits when one takes into account both a leptonic (or hadronic) jet and a dark matter spike in the inner part of M87. We further discuss the possibility of fitting the observed TeV $\gamma$-rays with the prompt emission from DM annihilations in the spike. We conclude in Sec.~\ref{conclusion}. Technical details can be found in the Appendices.

\section{Diffuse emission in the presence of a DM spike}
\label{DM_spike_only}

\begin{figure}[h]
\centering
\includegraphics[scale=0.45]{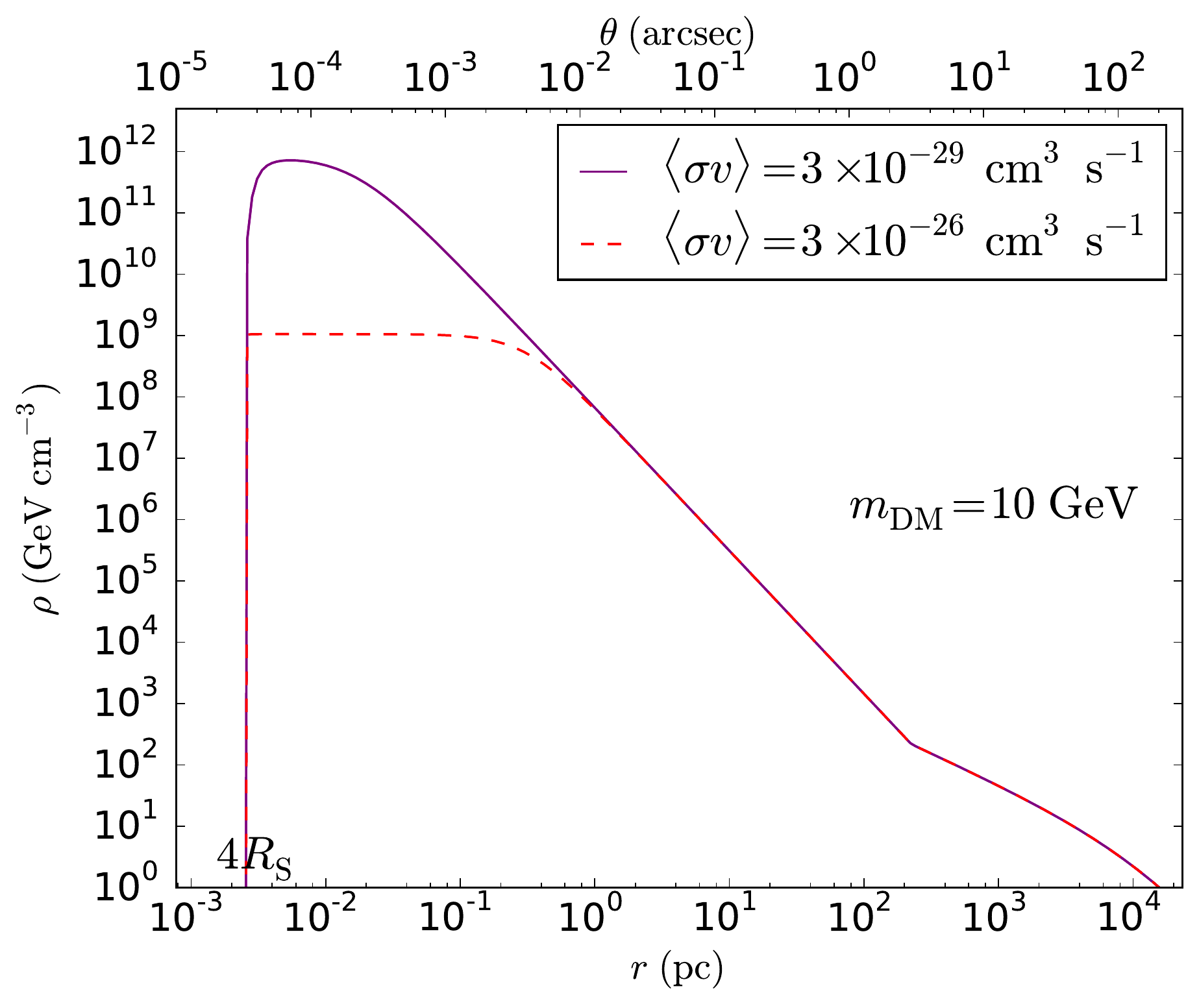}
\caption{\label{DM_profile}DM energy density as a function of the distance from the center, for a DM spike with $\gamma_{\mathrm{sp}} = 7/3$. DM annihilations soften the central cusp differently, depending on the DM mass and cross-section (as illustrated by the dashed and solid lines for a fixed DM mass and two different values of the annihilation cross-section).}
\end{figure}

The spike is modelled by a DM energy density $\rho \propto r^{-7/3}$, starting from the saturation radius, denoted by  $r_{\mathrm{ sat}}$ and determined by the DM mass and annihilation cross-section (see Fig.~\ref{DM_profile}), up to the spike radius  $r_{\mathrm{ sp}}$. Outside this inner region, i.e.~for $r > r_{\mathrm{sp}}$, we assume a NFW profile \cite{NFW}. More details are given in Appendix A. These assumptions are similar to those made in Ref.~\cite{Spike_GC_my_paper}---where we computed the boost in synchrotron radiation due to the presence of a spike in our galaxy---and to those of Ref.~\citep{Belikov2014} where the authors studied the enhancement of the diffuse extragalactic $\gamma$-ray background induced by DM spikes in other galaxies. In Fig.~\ref{DM_profile}, we show the resulting profiles for two different values of the DM annihilation cross-section, so as to illustrate the impact of DM annihilations on the value of the saturation radius. As one can see (and as is very well known), a larger cross-section tends to smooth down the profile.

The DM contribution to the spectral energy distribution (SED) of M87 is essentially three-fold. Photons can be produced in the observed regime by (i) intermediate and final-state radiation from the charged particles which are exchanged or produced in the DM annihilation process, (ii) decay or hadronization of the particles in the final state, and (iii) inverse Compton scattering (ICS) of electrons produced by DM annihilations off low energy photons (CMB, infrared and UV), as well as synchrotron emission from these DM-induced electrons. We will neglect any possible bremsstrahlung emission since M87 is deficient in cold gas \cite{Young2011}.

The first two emission modes (i and ii) are referred to as prompt emission. They are independent of the galaxy dynamics and are only determined by the DM particle physics properties. On the contrary, the third type of emission---namely synchrotron and ICS---strongly depends on the properties of the interstellar medium in the galaxy, such as the magnetic field strength and interstellar radiation field. In the presence of a very strong magnetic field, synchrotron emission becomes  the main source of low energy photons but it is also the main energy loss for the electrons and positrons produced by the DM. As a result, we find that prompt emission dominates ICS and is therefore the dominant source of high energy gamma-rays, while synchrotron radiation is the dominant source of X-rays.

The assumption of a very strong magnetic field in this work is justified by the presence of a BH at the center of M87. We will assume typically $B = 10^{10}-10^{11} \ \mu \rm G$, corresponding to the equipartition model, as discussed in Refs.~\cite{Regis&Ullio,Aloisio2004}. As a result, the electrons and positrons that are produced by the DM annihilations in the inner region are expected to stay confined to their site of injection, i.e. essentially in a sphere of radius $r_{\mathrm{sp}}$. This means that we can safely disregard spatial diffusion. 

We compute the prompt and synchrotron emission as follows. The prompt diffuse $\gamma$-ray intensity, $I_{\nu,i}^{\mathrm{prompt}}(\theta)$, at angle $\theta$ from the center, can be estimated by integrating the DM halo density over the line of sight (l.o.s.) coordinate $s$:
\begin{equation}
\label{I_nu_prompt}
\nu I_{\nu,i}^{\mathrm{prompt}}(\theta) \equiv E_{\gamma}^{2}\dfrac{\mathrm{d}n_{i}}{\mathrm{d}E_{\gamma}\mathrm{d}\Omega} = \dfrac{E_{\gamma}^{2}}{4\pi \eta} \dfrac{\left\langle \sigma v \right\rangle_{i}}{m_{\mathrm{DM}}^{2}} \dfrac{\mathrm{d}N_{\gamma,i}}{\mathrm{d}E_{\gamma}} \int_{\mathrm{l.o.s.}} \! \rho^{2}\left( r(s,\theta)\right)  \, \mathrm{d}s
\end{equation}
where $\eta$ (= 2 here) is a factor that accounts for the nature (real/complex, Majorana/Dirac) of DM, $\mathrm{d}N_{\gamma,i}/\mathrm{d}E_{\gamma}$ is the $\gamma$-ray spectrum that originates from the specific annihilation channel $i$ (see Ref.~\cite{Cirelli_cookbook}) and $\left\langle \sigma v \right\rangle_i$ the DM annihilation 
cross-section into this final state. Finally we use $E_{\gamma} = h \nu$ with $h$ the Planck constant.

To compute the synchrotron intensity, we need one additional step. We first need to determine the electron and positron spectrum from the DM annihilation rate, using 
\begin{equation}
\label{psie}
\psi_{\mathrm{e},i}(r,E) = \dfrac{1}{b(r,E)} \dfrac{\left\langle \sigma v \right\rangle_{i}}{\eta} \left( \dfrac{\rho(r)}{m_{\mathrm{DM}}} \right) ^{2} \int_{E}^{m_{\mathrm{DM}}} \ \dfrac{\mathrm{d}N_{\mathrm{e},i}}{\mathrm{d}E_{\mathrm{S}}}  \ \mathrm{d}E_{\mathrm{S}},  
\end{equation}
and convolve it with the synchrotron power, $P_{\nu}(r,E)$ (see Appendix B), so as to obtain  the synchrotron emissivity 
\begin{equation}
j_{\nu,i}(r) = 2 \int_{m_{\mathrm{e}}}^{m_{\mathrm{DM}}} \ P_{\nu}(r,E) \psi _{\mathrm{e},i}(r,E) \, \mathrm{d}E. 
\label{jnu}
\end{equation}
The term $\mathrm{d}N_{\mathrm{e},i}/\mathrm{d}E_{\mathrm{S}} $ in Eq.~\eqref{psie} represents the number of electrons\footnote{ $\mathrm{d}N_{\mathrm{e},i}/\mathrm{d}E_{\mathrm{S}} \sim \delta(E_{\mathrm{S}} - m_{\rm{DM}})$ if the DM directly annihilates into electrons and positrons.} produced by the decay or hadronization of the final state $i$. The total emissivity is the sum of the electron and positron contributions, hence the factor 2 in Eq.~\eqref{jnu}, to account for the fact that a positron is always produced simultaneously with an electron. 

The term $b(r,E)$ corresponds to the total energy loss rate. Given the large values of the magnetic field that we consider in this paper, the losses are dominated by the synchrotron losses, that is 
\begin{equation}
b(r,E) \equiv b_{\mathrm{syn}}(r,E) = \dfrac{4}{3} \sigma_{\mathrm{T}} c \dfrac{B(r)^{2}}{2 \mu_{0}} \, \gamma_{\mathrm{L}} ^{2},
\end{equation}
where $ \sigma_{\mathrm{T}} $ is the Thomson cross-section, $ B(r) $ the intensity of the magnetic field, $c$ the speed of light, $\gamma_{\mathrm{L}} = E/(m_{\mathrm{e}} c^{2})$ the Lorentz factor of the electrons, $m_{\mathrm{e}}$ the electron mass and $ \mu_{0} $ the vacuum permeability, see e.g.~Ref.~\cite{Longair2011}. The synchrotron intensity finally reads
\begin{equation}
\label{I_nu_syn}
\nu I_{\nu,i}^{\mathrm{syn}}(\theta) = \nu \int_\mathrm{l.o.s.} \! \dfrac{j_{\nu,i}(r(s,\theta))}{4 \pi} \, \mathrm{d}s.
\end{equation}
Note that both injection spectra $\mathrm{d}N_{\gamma,i}/\mathrm{d}E_{\gamma}$ and $\mathrm{d}N_{\mathrm{e},i}/\mathrm{d}E_{\mathrm{S}}$ in Eqs.~\eqref{I_nu_prompt} and \eqref{psie} are taken from Ref.~\cite{Cirelli_cookbook} and include electroweak corrections as these become increasingly important at high energy.
Also, to perform the line-of-sight integration, we will assume that the magnetic field distribution in the inner region is shaped by the accretion flow. Hence, unless stated otherwise, we will assume that the radial dependence of the magnetic field is given by the equipartition model and thus goes as $ B(r) \propto r^{-5/4}$, as discussed in Refs.~\cite{Regis&Ullio,Aloisio2004}. This specific form actually leads to very large values of the magnetic field toward the center, typically up to $10^{10}-10^{11}\ \rm \mu G$ in the very inner region---as mentioned previously---which is at least eight orders of magnitude larger than the values usually considered in the Milky Way (see Appendix B).

In principle, one should also take into account the effect of advection of electrons and positrons toward the center by the accretion flow around the BH, which increases the synchrotron flux typically in the range $10^{12} - 10^{14}\ \rm Hz$ \cite{Aloisio2004}. However, we disregard this effect throughout the present  paper since most of our constraints come from higher frequencies, given the large magnetic field strengths we consider. Moreover, even for smaller magnetic fields, including advection would not weaken our constraints but would only make them more stringent, so we remain conservative in this regard.

\section{Upper limits on the annihilation cross-section}
\label{DM_upperlimits}

Limits are set on the DM annihilation cross-section by comparing the expected emission from DM with the measured SED for M87. Most data points have actually been compiled by the Fermi Collaboration in Ref.~\citep{Fermi_M87}. We use in particular:
\begin{itemize}
\item the historical measurements of the core emission (from millimeter  to  X-rays \cite{Biretta1991,Despringre1996,Tan2008,Shi2007,Perlman2001,Sparks1996,Marshall2002}),  
\item  the MOJAVE VLBA data point at 15 GHz which was derived in Ref.~\citep{Fermi_M87} (the data was reported in Ref.~\cite{Lister2009}), 
\item the 2009 X-ray data points which were derived in Ref.~\citep{Fermi_M87} from the 2009 Chandra measurements \citep{Harris2009}, 
\item the 2009 Fermi-LAT data \citep{Fermi_M87}, 
\item the 2004 HESS data \citep{Aharonian2006}, 
\item the 2007 VERITAS data \citep{Acciari2008}, 
\item the 2011 MAGIC data \citep{MAGIC}, 
\end{itemize}
which essentially give us the observed value of the electromagnetic flux between $10^{10}$ and $10^{27}$ Hz.

\begin{figure*}[t]
\centering
\includegraphics[scale=0.43]{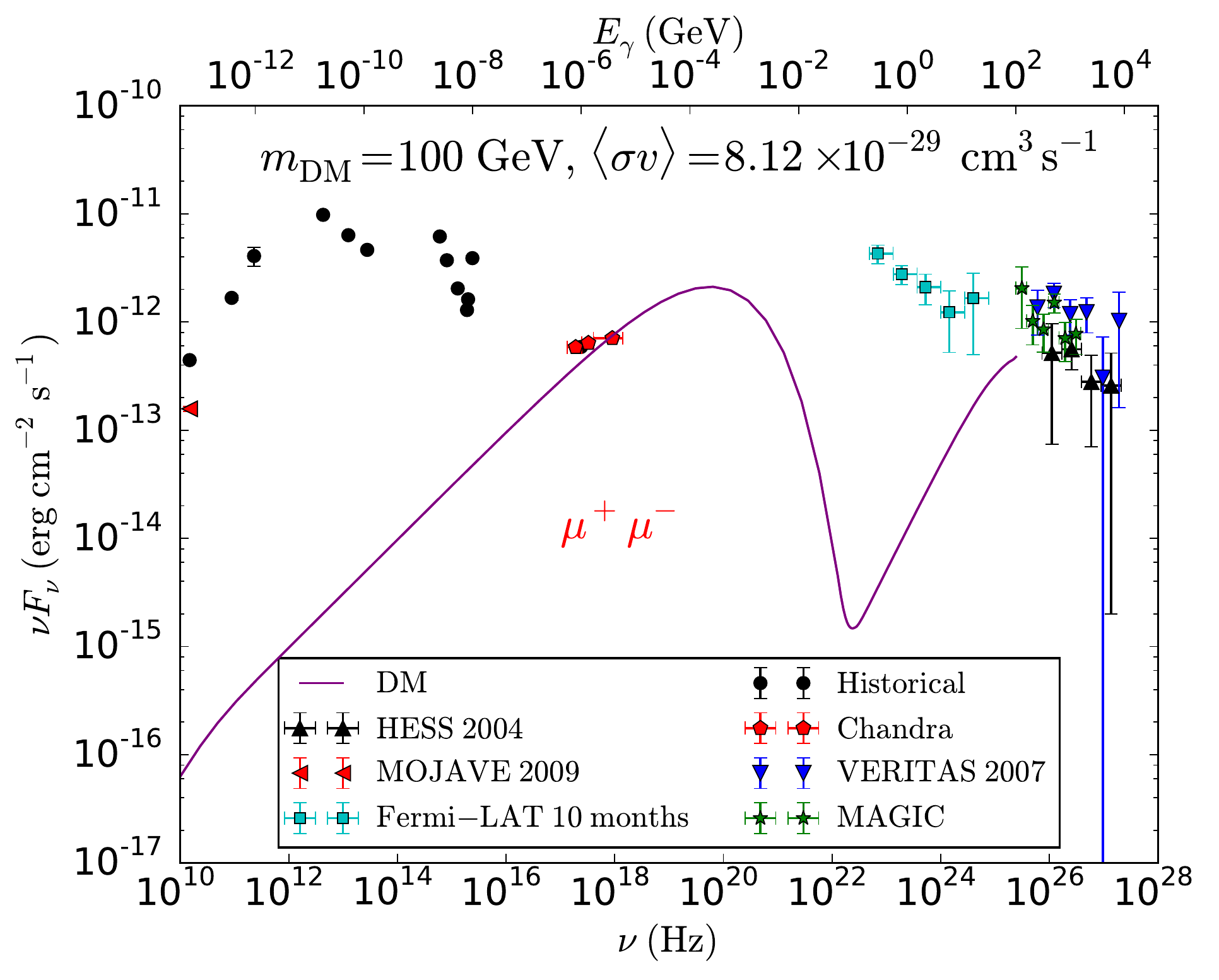}\includegraphics[scale=0.43]{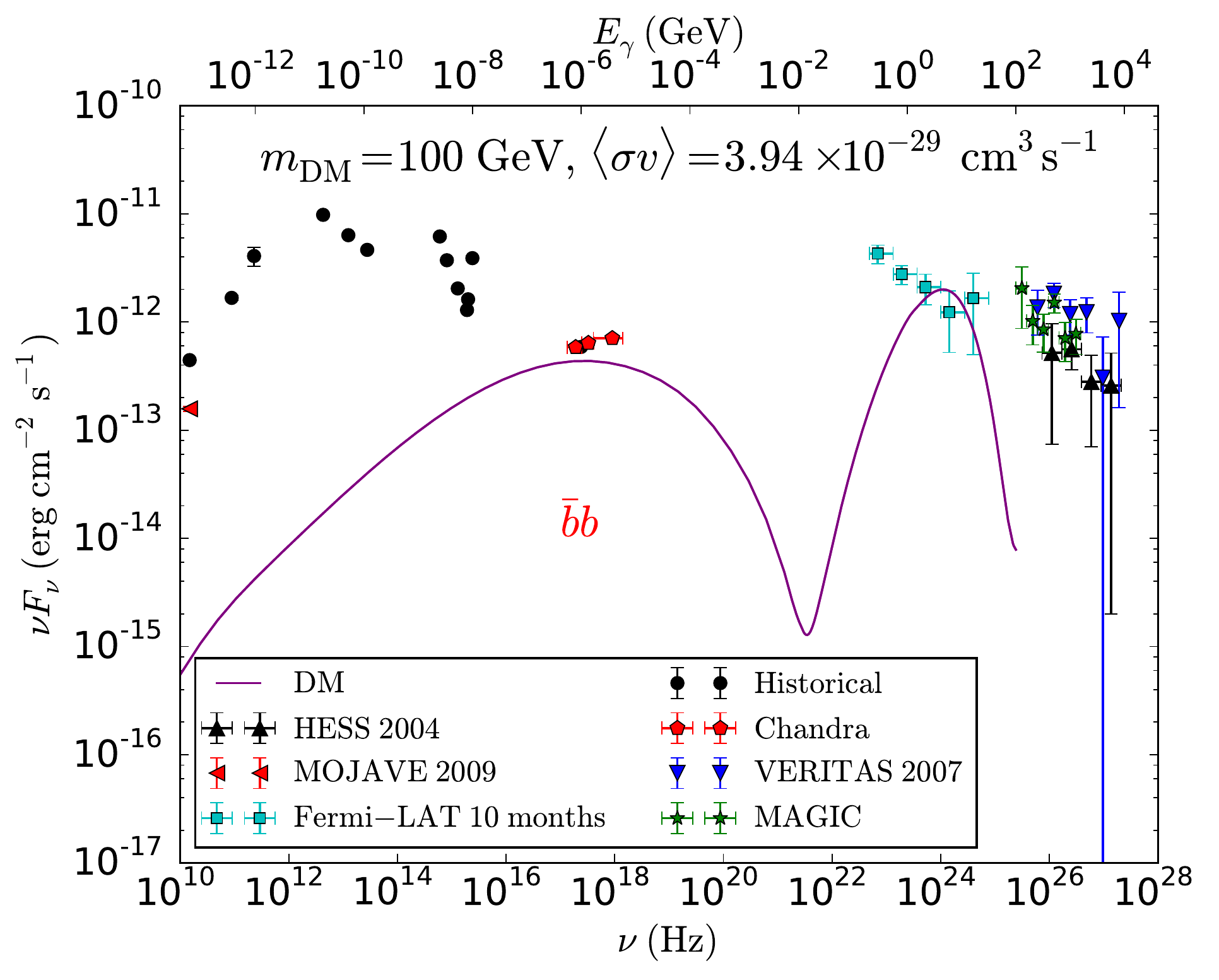}
\caption{\label{spectrum_DM_only}SED of M87 from the millimeter band to TeV $\gamma$-rays, from a DM spike, for 100 GeV DM particles annihilating into $\mu^+\mu^-$ with $\left\langle \sigma v \right\rangle = 8.12 \times 10^{-29}\ \rm cm^{3}\ s^{-1}$ (left panel), and into $\bar{b}b$ with $\left\langle \sigma v \right\rangle = 3.94 \times 10^{-29}\ \rm cm^{3}\ s^{-1}$ (right panel). These are the maximal values of the cross-section compatible with the data for a DM candidate of 100 GeV.}
\end{figure*}

\subsection{Methodology}

The dark matter contribution  is estimated by integrating the prompt and synchrotron intensities $\nu I_{\nu,i}^{\mathrm{prompt},\mathrm{syn}}$ [see Eqs.~\eqref{I_nu_prompt} and \eqref{I_nu_syn}] over a field of view that is centred on the galactic center and set by the angular resolution of the relevant experiment for a given frequency. Given the spherical symmetry of the spike, the prompt and synchrotron spectra are given by 
\begin{equation}
\label{iresolution}
\nu F_{\nu,i}^{\mathrm{prompt},\mathrm{syn}} = 2 \pi \int_0^{\theta_{\mathrm{res}}} \! \nu I_{\nu,i}^{\mathrm{prompt},\mathrm{syn}}(\theta) \sin \theta \, \mathrm{d}\theta.
\end{equation}
We recall that Chandra has an angular resolution of 0.5 arcsec over the whole energy range considered here \citep{Weisskopf2002} while the angular resolution of the Fermi experiment reads $0.8^\circ \times E_{\mathrm{GeV}}^{-0.8}$ \citep{Fermi_M87} ($E_{\mathrm{GeV}}$ is the energy of the $\gamma$-rays normalized to 1 GeV). HESS, VERITAS and MAGIC all have angular resolutions of the order of $0.1^\circ$, see Refs~ \cite{Benbow2005,Holder2011,Colin2009}. In practice though, the spike is contained in such a small region that the exact value of the upper bound of the integral in Eq.~\eqref{iresolution} does not significantly affect the result.

To set limits, we require that the synchrotron and prompt emission fluxes that are induced by the DM annihilations do not exceed the error bars on the flux for any measured data point. More specifically, we exclude any value of the annihilation cross-section that satisfies the following inequality for any observed frequency $\nu$:
\begin{equation}
S_{\nu}^{\mathrm{model}} - (S_{\nu}^{\mathrm{obs}} + \Delta S_{\nu}^{\mathrm{obs}}) \geq \ \kappa \ (S_{\nu}^{\mathrm{obs}} + \Delta S_{\nu}^{\mathrm{obs}}),
\end{equation}
with $S_{\nu} \equiv \nu F_{\nu}$ and $\kappa \ll 1$ (typically $\kappa = 10^{-4}$).  The terms $S_{\nu}^{\mathrm{model}}$, $S_{\nu}^{\mathrm{obs}}$ and $\Delta S_{\nu}^{\mathrm{obs}}$ represent respectively the expected DM contribution, the observed SED, and the $1\sigma$ error bar at frequency $\nu$. 

Note that large values of the annihilation cross-section, i.e.
\begin{equation}
\left\langle \sigma v \right\rangle \gg 10^{-27} \left( m_{\mathrm{DM}}/(10\ \mathrm{GeV})\right) \ \rm cm^3\ s^{-1},
\end{equation}
flatten the inner part of the spike below a saturation radius given by 
\begin{equation}
r_{\mathrm{sat}} \sim 4 \times 10^{-2} \left( \frac{\left\langle \sigma v \right\rangle}{10^{-27}\ \rm cm^3\ s^{-1}}\right) ^{1/2} \left( \frac{m_{\mathrm{DM}}}{10\ \mathrm{GeV}}\right) ^{-1/2}\ \rm pc.
\end{equation}
Therefore one cannot always rescale the flux for different values of $\left\langle \sigma v \right\rangle$, since in some cases the cross-section actually modifies the DM profile. For $\left\langle \sigma v \right\rangle \lesssim 10^{-27} \left( m_{\mathrm{DM}}/(10\ \mathrm{GeV})\right)\ \rm cm^3\ s^{-1}$, on the other hand, the saturation radius is very small (below $10^{-2}\ \rm pc$) and in this regime the fluxes that we compute are simply proportional to the annihilation cross-section.

\subsection{Results}

\begin{figure*}[t]
\centering
\includegraphics[scale=0.45]{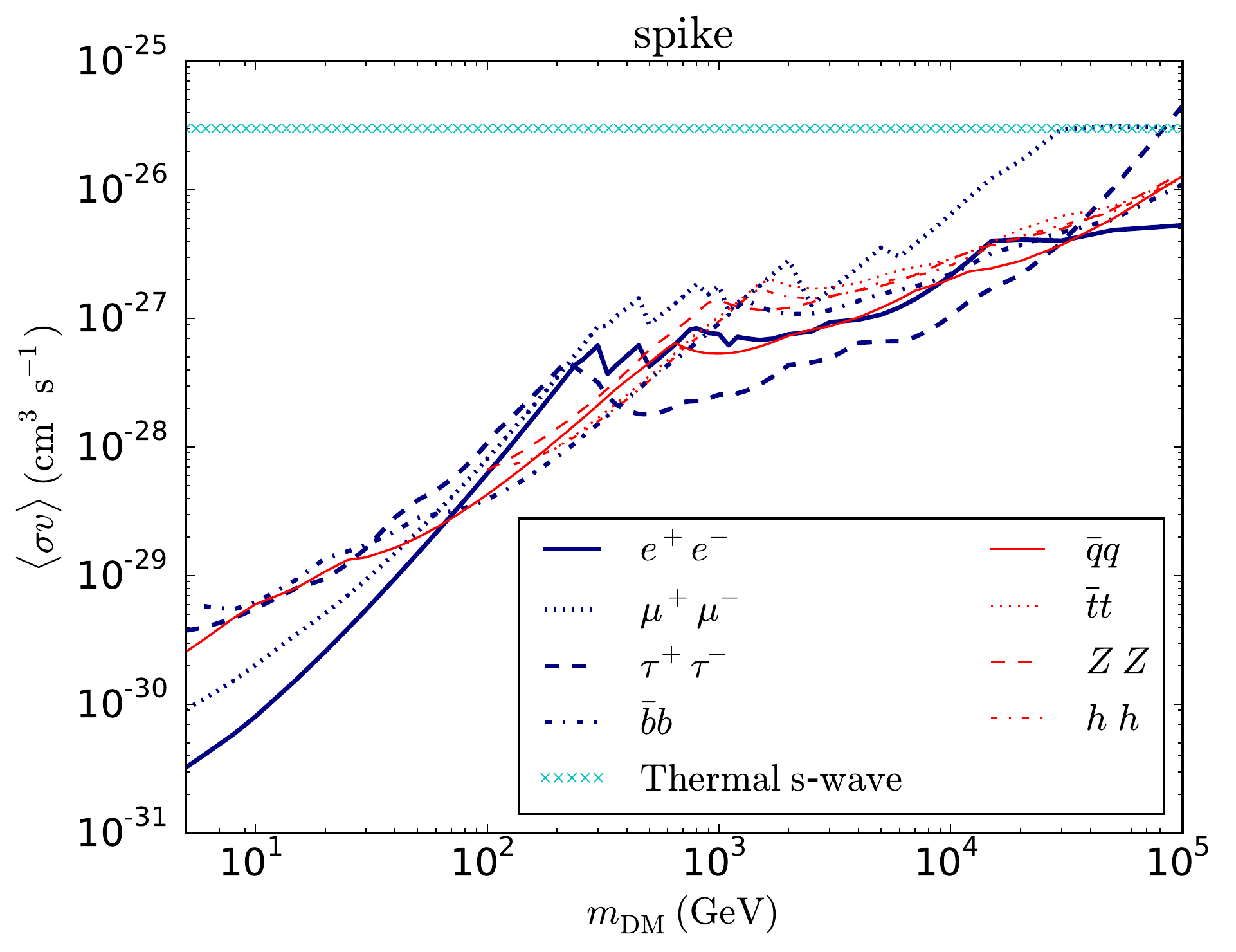}\includegraphics[scale=0.45]{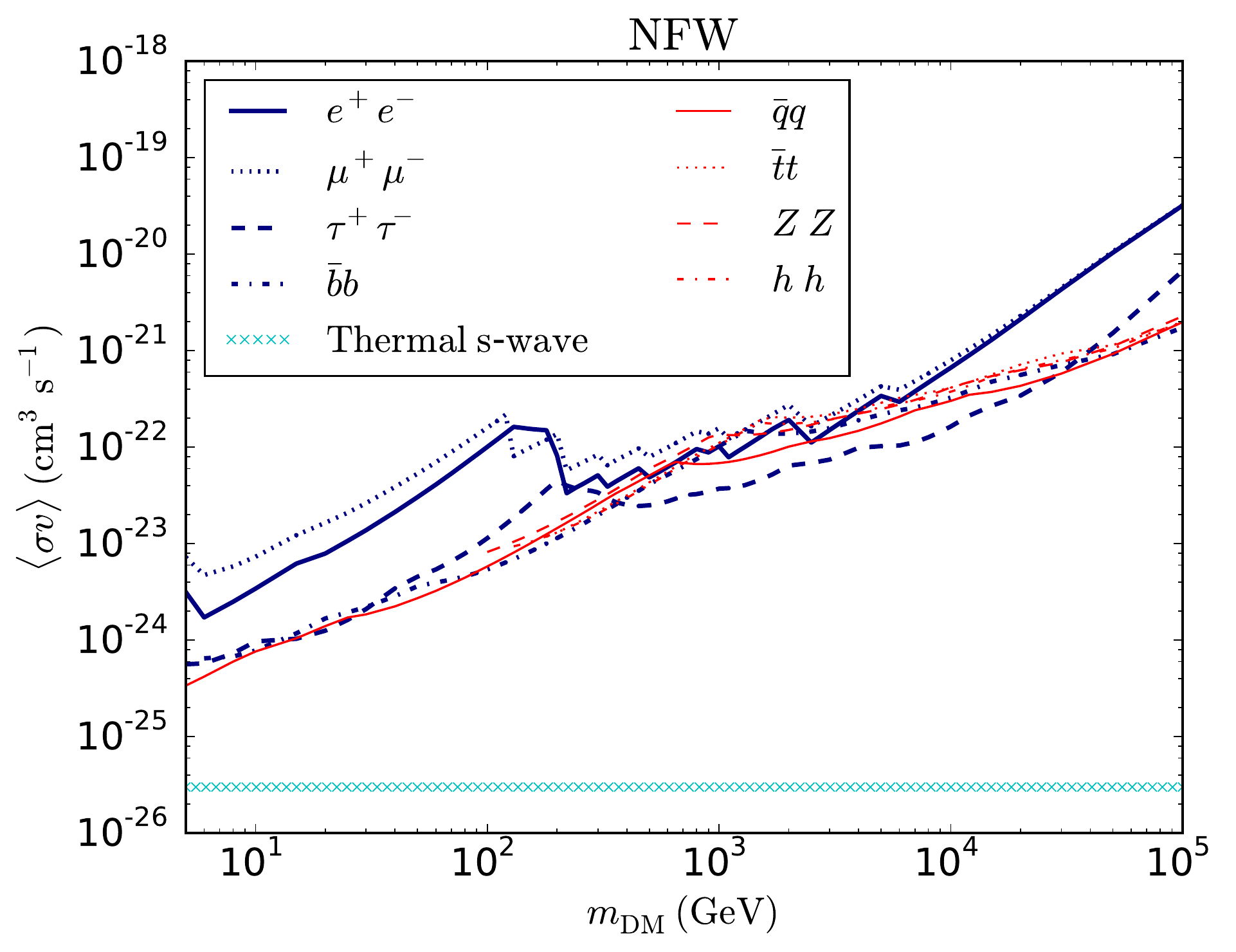}
\caption{\label{limits}Upper limits on the annihilation cross-section as a function of DM mass for various annihilation channels. Constraints derived assuming a spike in the DM distribution are shown in the left panel while in the right panel we show the constraints obtained assuming a standard NFW cusp.}
\end{figure*}

In Fig.~\ref{spectrum_DM_only}, we plot the largest allowed electromagnetic emission (prompt plus synchrotron)  expected from DM annihilations for a 100 GeV DM candidate. The left panel shows the predictions for DM annihilations into $\mu^+\mu^-$ while the right panel shows the predictions for annihilations into $\bar{b}b$.  The two bumps correspond to the synchrotron (left) and prompt (right) emission.  We also derive the constraints on the annihilation cross-section for any DM mass and eight annihilation channels ($e^+e^-$, $\mu^+\mu^-$, $\tau^+\tau^-$, $\bar{q}q$, $\bar{b}b$, $\bar{t}t$, $Z Z$, $h h$, with $h$ the standard model Higgs boson and $q=u,d,s$), shown in Fig.~\ref{limits}. The left panel shows the constraints in the presence of a spike and the right panel shows the constraints without a spike (assuming a NFW profile).

The constraints in the case of a spike essentially rule out any DM candidate with a thermal (s-wave) cross-section, from a few GeV to a hundred TeV.\footnote{These constraints can be extended down to the MeV range for leptons and light quarks, provided the magnetic field is of the order of the equipartition value. Synchrotron emission then peaks around $10^{10}$-$10^{11}\ \rm GHz$. For smaller values of the magnetic field, the synchrotron peak falls below $10^{10}\ \rm GHz$, i.e.~in the radio range, where synchrotron self-absorption significantly reduces the flux (see e.g. Ref.~\citep{Aloisio2004}), thus preventing one from setting any constraints on the DM annihilation cross-section.} In other words, they rule out the entire mass range relevant for thermal DM.  The only exceptions to our generic conclusions are for candidates which mostly annihilate into $\mu^+ \mu^-$ (the limit is then about 30 TeV) or which annihilate democratically into all channels (but the limit would still be close to 100 TeV nevertheless).  We also confirm that thermal candidates with a p-wave suppressed annihilation cross-section are ruled out if they are much lighter than a few GeV. For comparison, our limits in the case of a pure NFW profile are considerably weaker, see Fig.~\ref{limits} (right panel). They only rule out very light (below a few GeV) s-wave thermal DM candidates. 

Of course our conclusions rely on the crucial assumption of the existence of a spike with $\gamma_{\mathrm{sp}} \sim 7/3$. While this remains speculative, the stability of such a spike with respect to  the scattering off stars is very likely. Hence, if the initial conditions were such that a spike could form in M87, our constraints would rule out a very large chunk of the thermal DM parameter space. An alternative interpretation of our results is that the discovery of a thermal s-wave DM candidate would rule out the existence of a spike in M87. This would in turn constrain the evolution and  formation of the supermassive BH at the center of M87.

\subsection{Robustness of our constraints}

In the previous sections, we have considered extremely large values of the magnetic field (several orders of magnitude with respect to the Milky Way) and we have neglected absorption. It is therefore legitimate to question the robustness of our limits with respect to the magnetic field model and absorption processes.

\subsubsection{Dependence on the magnetic field}

Our most stringent limit on the DM contribution in M87 is set by the Chandra X-ray data. Since most of the DM induced X-ray signal originates from synchrotron radiation and synchrotron emission strongly depends on the magnetic field, a weaker magnetic field could weaken our constraints. This is particularly worrying for the  $e^+e^-$ and $\mu^+\mu^-$ final states which give the largest X-ray contribution in M87 when $m_{\mathrm{DM}} \lesssim 100\ \rm GeV$, as shown in Fig.~\ref{spectrum_DM_only}, and which could  become unconstrained. 
 
However a weaker magnetic field in the inner region, taken for example to be constant and about $B = 10^{5}\ \mu \rm G$ (as suggested in Ref.~\cite{Neronov2007}), only weakens our constraints by about one order of magnitude below 30 GeV. We thus get $\left\langle \sigma v \right\rangle  < 10^{-29}\ \rm cm^3\ s^{-1}$ for $e^+e^-$ and $\mu^+\mu^-$ instead of $\left\langle \sigma v \right\rangle < 10^{-30}\ \rm cm^3\ s^{-1}$. Hence, even in the case of a weaker magnetic field in the inner region, we can rule out thermal s-wave DM. Note that decreasing the magnetic field to $B = 1 \rm{mG}$ makes our constraints  stronger again as the signal would be constrained by the MOJAVE data. Finally, as noted in Ref.~\citep{Aloisio2004}, if the magnetic field is significantly smaller than the equipartition value, synchrotron self-Compton emission decreases the DM-induced electron spectrum and thus also the synchrotron flux accordingly. Based on the results of Ref.~\citep{Aloisio2004}, we estimate that our limits for the $e^+e^-$ and $\mu^+\mu^-$ channels are weakened by an additional order of magnitude below  $\sim 100\ \rm GeV$ for a magnetic field strength weaker than $10^{5}\ \mu \rm G$. However this does not affect our conclusion since we can still exclude thermal s-wave DM.
  
A magnetic field of the order of $10^{5}\ \mu \rm G$ also changes the limits for candidates with a mass above $\sim 50$ TeV for the $e^+e^-$ and $\mu^+\mu^-$ channels. Indeed, the synchrotron peak then falls in the energy range [100 keV, 100 MeV] where there are no data. In that case, the limit is given by the prompt component and weakened to the level of $3 \times 10^{-25}\ \rm cm^{3} \ s^{-1}$. As a reminder,  in the case of a stronger magnetic field, we could rule out the canonical thermal cross-section for the $e^+e^-$ channel. Thus, for such a value of the magnetic field, one can no longer exclude s-wave DM for particles heavier than 50 TeV. It is worth pointing out though that a smaller value of the magnetic field of e.g.~$10^{3}\ \mu \rm G$ would not alleviate our constraints as it would give rise to an excess in X-rays, which has not been observed. Such a value would therefore lead to an exclusion limit instead, similar to the one obtained for the equipartition magnetic field.

Finally, the annihilation channels which give a softer electron spectrum, e.g.~the $\bar{b}b$ channel, are unaffected by a weaker magnetic field since the prompt $\gamma$-ray emission dominates the synchrotron emission, see Fig.~\ref{spectrum_DM_only}, and therefore our limits for these channels are independent of the magnetic field.

\subsubsection{Absorption}

Absorption is another process which could weaken our conclusions. We may have overestimated the flux by not accounting for the photons which have been emitted by synchrotron radiation and absorbed by the same electron population that produced them. Since the authors of \citep{Aloisio2004,Regis&Ullio} showed that this effect is only very efficient below $10^{10}\ \rm Hz$, we cut the synchrotron emission below this critical frequency. This prevents us from constraining the scenarios emitting in this energy range, i.e.~candidates lighter than $O(1)$ GeV, unless the magnetic field is strong.

At the other end of the spectrum, high-energy $\gamma$-rays could also be absorbed via $e^+ e^-$ pair production with the background radiation field. However, the authors of Refs.~\citep{Neronov2007,Brodatzki2011}  showed that the inner region of M87 is transparent to $\gamma$-rays below $\sim$ 10 TeV. Since we only have data below 10 TeV and absorption is relevant only above 10 TeV, we can neglect this effect for all candidates below 10 TeV. For the much heavier candidates, the lack of data above 10 TeV  makes absorption irrelevant for the moment. Hence we have neglected absorption in our study.

\section{DM spike and jet}
\label{spike+jet}

In the previous section, we have investigated the DM contribution to the SED of M87 but neglected the contribution from the BH. In reality the jet emission associated with the BH must be taken into account. Indeed, to be observable, any putative emission from DM should be brighter than the emission from the jet.

\subsection{Jet emission}

\begin{figure}[h!]
\centering
\includegraphics[scale=0.45]{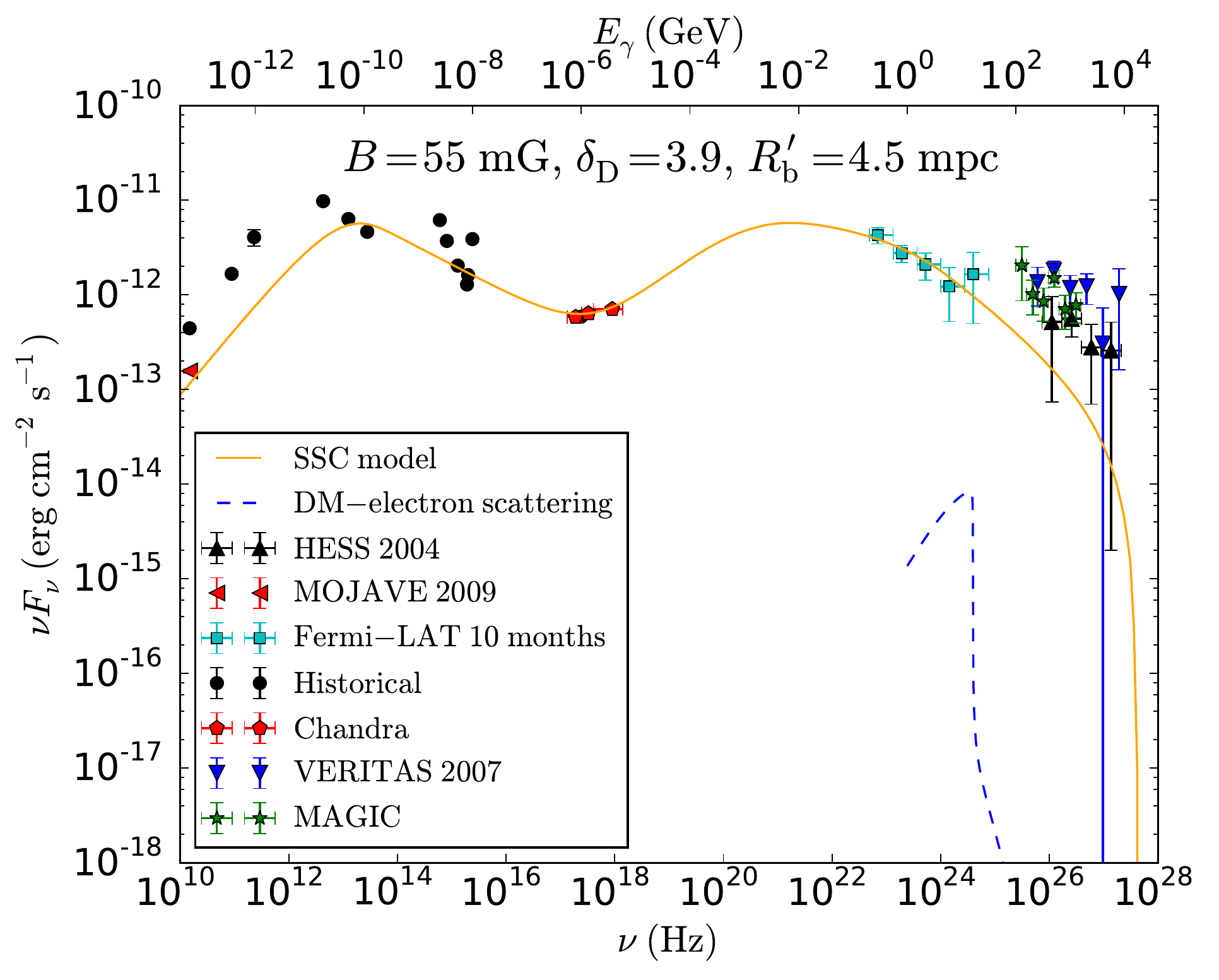}
\caption{\label{SED_jet}SED of M87 from the millimeter band to TeV $\gamma$-rays. The orange solid line represents the SSC model that best fits the data, for a 55 mG magnetic field in the plasma blob, a Doppler factor of 3.9, and a radius of 4.5 mpc for the blob in its rest frame, as found in Ref.~\citep{Fermi_M87}. The blue dashed line corresponds to the signal expected from DM-jet scattering, as described in Ref.~\citep{Gorchtein_DM_AGN_jet}. Details on the parameters can be found in the text.}
\end{figure}

The mechanism giving rise to high energy photons from the jet is not well known. It is unclear whether these photons have a leptonic or hadronic origin. In the most popular model, the $\gamma$-rays originate from electrons contained in a blob of plasma that moves relativistically and possesses a fairly strong magnetic field. This leptonic synchrotron self-Compton emission (SSC) proceeds in two steps: the electrons from the jet first produce photons in the infrared band due to synchrotron radiation in a strong magnetic field. Then, in a second step, these low energy photons are upscattered to $\gamma$-ray energies by ICS on the same electron population that produced them. 

To take this effect into account and fit the spectral energy distribution of M87, we  use the model described in Ref.~\citep{SSC_model} and take the best-fit SSC parameters given in Ref.~\cite{Fermi_M87}, see Appendix C.  The parameters we consider are: a Doppler factor $\delta_{\mathrm{D}} = 3.9$, a magnetic field $B = 55\ \rm mG$, and a source radius $R'_{\mathrm{b}} = 4.5\ \rm mpc$ in the rest frame of the blob. The data that have been used for the fit are the 2009 MOJAVE, Chandra and Fermi-LAT data. The best-fit value for the normalisation of the electron distribution is $K = 5.81 \times 10^{51}$. The corresponding SSC emission for this set of parameters is shown in Fig.~\ref{SED_jet} (see the orange solid line).

Note that the scattering of the DM particles off electrons and protons in the jet might also produce high energy photons \citep{Gorchtein_DM_AGN_jet} and, consequently, lead to a characteristic signature in the Fermi-LAT data. The associated flux is proportional to the integral of the DM density over the line of sight ($\delta_{\mathrm{DM}}$, as in \citep{Gorchtein_DM_AGN_jet}), and the jet power ($L$). For M87, the highest allowed jet power is $L \sim 10^{45}\ \rm erg\ s^{-1}$ \citep{Fermi_M87}. For an optimal configuration of the DM spike, i.e. for the largest possible DM energy density (correspondingly to the smallest possible saturation radius, typically $\left\langle \sigma v \right\rangle \sim 10^{-30}-10^{-29} \ \rm{cm^3\ s^{-1}}$ for $m_{\rm{DM}} \sim 100$ GeV), the line-of-sight integration gives $\delta_{\mathrm{DM}} \sim 10^{9}\ \rm M_{\odot}\ pc^{-2}$ which leads to a $\gamma$-ray flux of $\sim 10^{-14}\ \rm erg\ cm^{-2}\ s^{-1}$. This is roughly two orders of magnitude below the Fermi data, as shown in Fig.~\ref{SED_jet} (blue dashed line). This process is therefore subdominant for M87, and we will disregard it in the following discussion.

\begin{figure}[h!]
\centering
\includegraphics[scale=0.45]{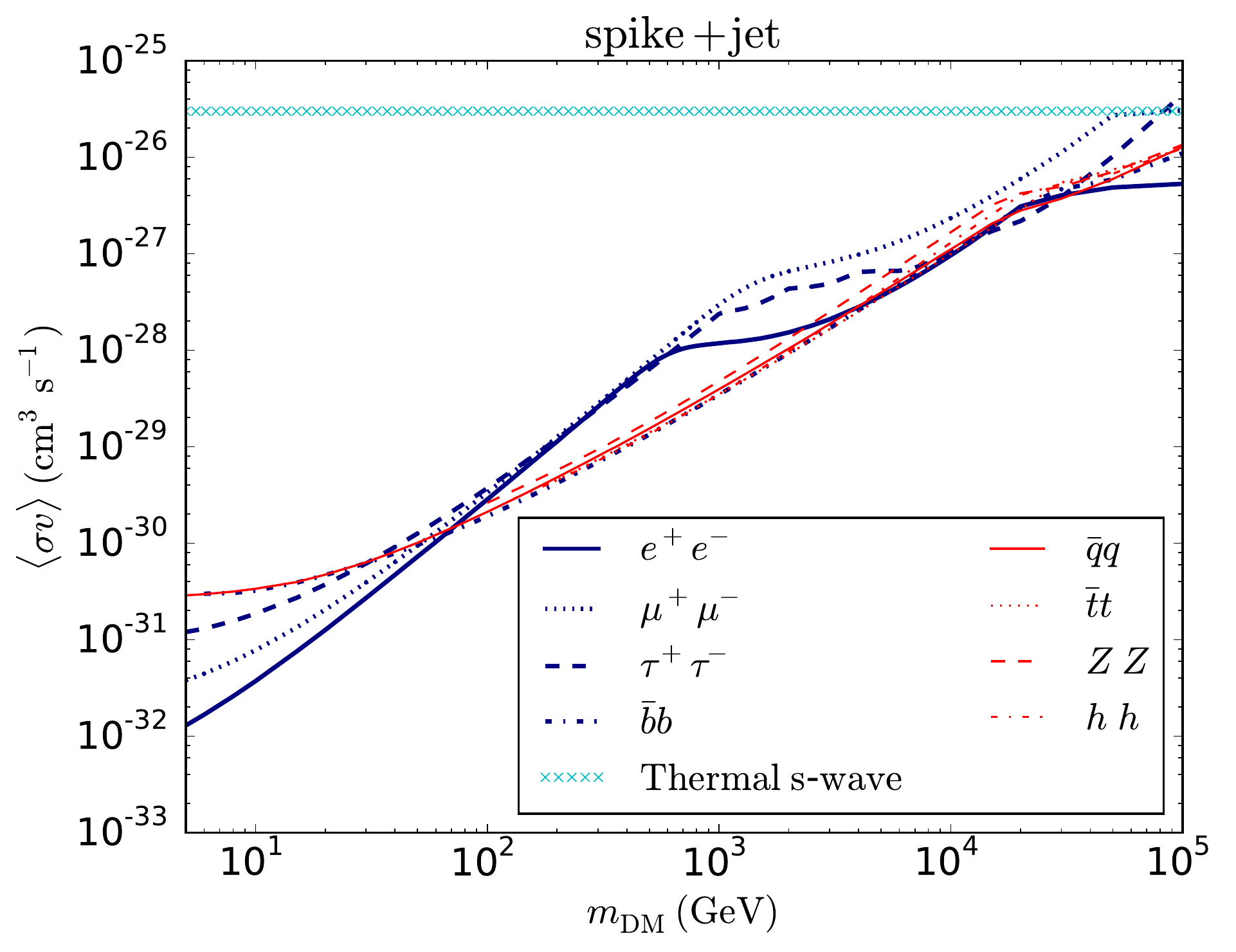}
\caption{\label{limits_with_jet}Upper limits on the annihilation cross-section as a function of DM mass for usual annihilation channels, obtained after summing the DM contribution to the photon emission expected from the jet (using the SSC model) and excluding cross-sections that depart from the best fit at $2\sigma$.}
\end{figure}

\begin{figure*}[t]
\centering
\includegraphics[scale=0.45]{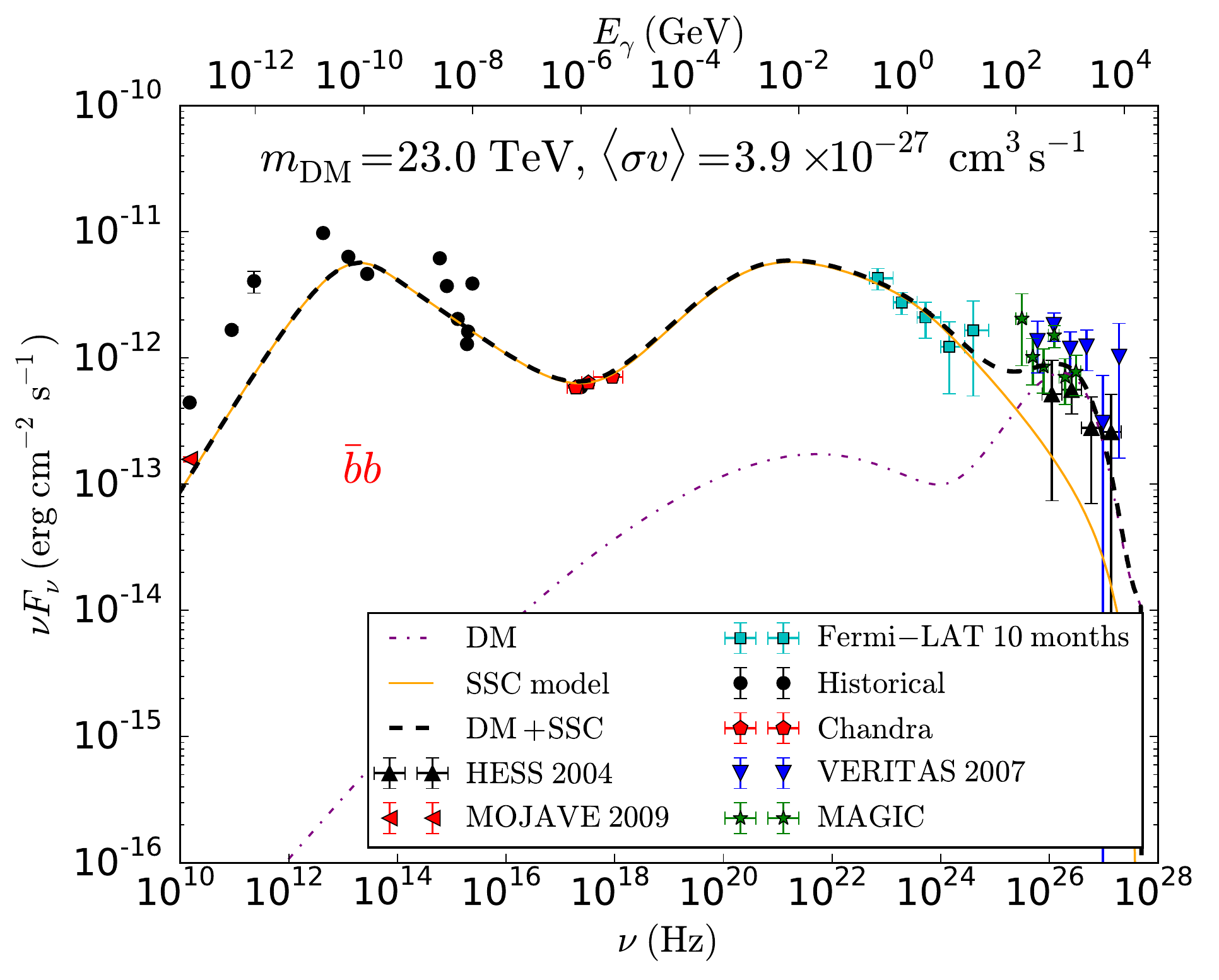}\includegraphics[scale=0.45]{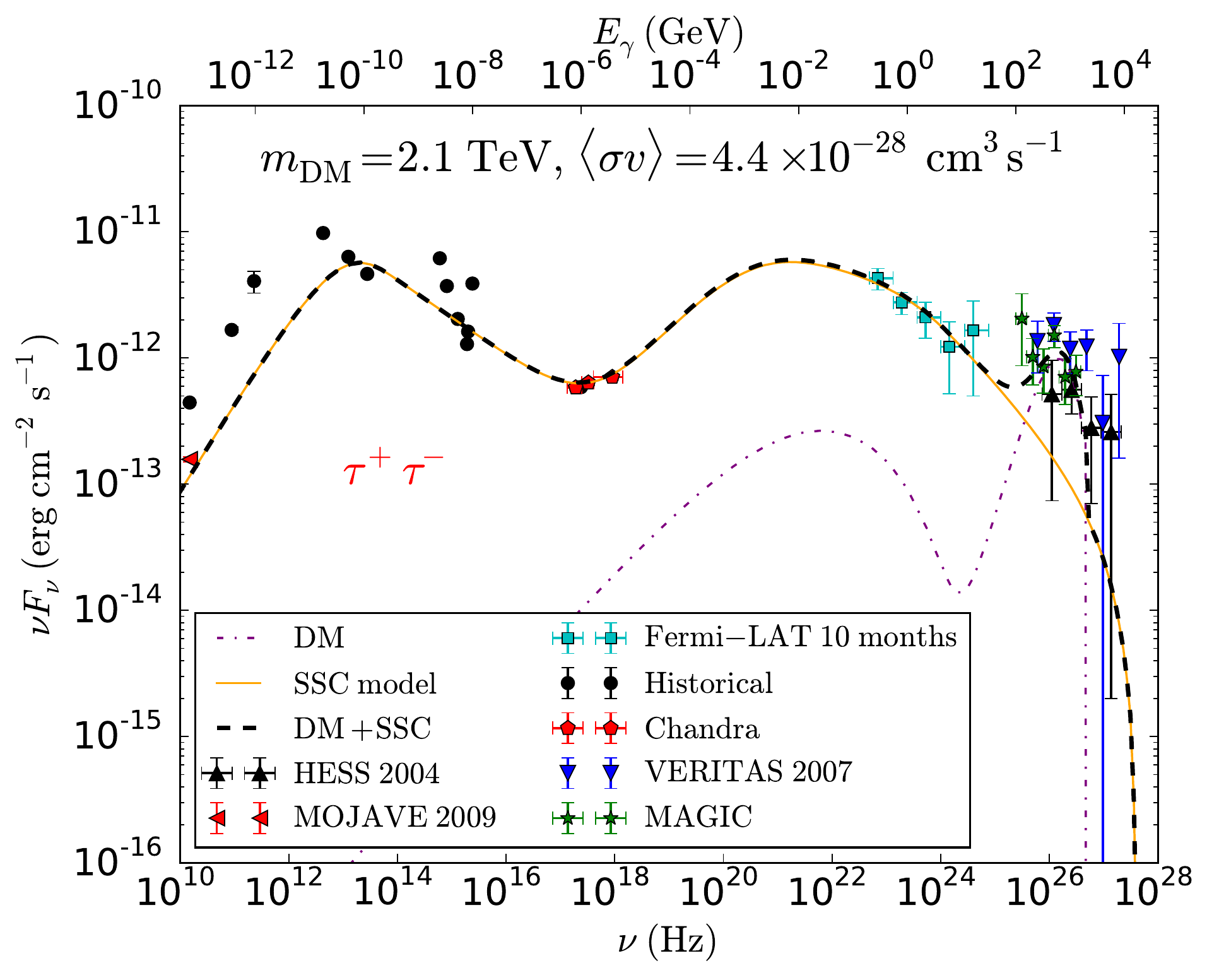}
\caption{\label{SSC+DM}SED of M87 from the millimeter band to TeV $\gamma$-rays. The SSC model for the jet gives a double peak structure (orange solid line). The contribution from the DM spike is depicted by the purple dot-dashed line, for annihilations into $\bar{b}b$ (left panel) and $\tau^+\tau^-$ (right panel), with the synchrotron peak around $10^{22}\ \rm Hz$ and the prompt emission peak around $10^{26}\ \rm Hz$. The black dashed line is the total SED.}
\end{figure*}

\subsection{Upper limits on the annihilation cross-section with spike+jet}

Because the jet emission associated with the SSC model fits the data very well up to $\gamma$-ray energies of 100 GeV, there is little room for a dark matter contribution to the SED of M87 for candidates lighter than 20 TeV. Our results are shown in Fig.~\ref{limits_with_jet}. We have excluded any values of the annihilation cross-section that cause the total flux to depart from the best fit to the Chandra and Fermi data by more than $2 \sigma$ and improved the limits by about one order of magnitude with respect to the constraints derived without considering the jet emission. Above  $E_{\gamma} \gtrsim$ 100 GeV,  a DM contribution improves the fit and prevents us from setting a better limit. Note that we can also exclude thermal p-wave DM up to $\sim 40$ GeV. 

Because the jet emission fits the whole spectrum up to 100 GeV and, in particular, fits the Chandra data which have the smallest error bars, any additional DM contribution---even small---tends to worsen the chi-square statistic and thus leads to stronger constraints. Since the Chandra data constrain the synchrotron contribution which is very sensitive to the magnetic field, these constraints strongly depend on the strength of the magnetic field.

Finally, let us recall that the limits derived in this section depend on the underlying jet model which, as we mentioned, is still very uncertain. Yet, the limits of Fig.~\ref{limits_with_jet} do illustrate the importance of including a model for the jet.

\subsection{Explaining the TeV data with a DM spike}

\subsubsection{Fits with a jet + DM spike}

As shown in Fig.~\ref{SSC+DM}, the simplest SSC model does not explain the TeV emission measured by HESS, MAGIC and VERITAS (although some of the points are consistent with the jet model).\footnote{Also according to the authors of Ref.~\citep{deJong2015}, it may be possible to refine the SSC model so as to fit the TeV data.} This led the authors of Ref.~\cite{Saxena2011} to discuss the possibility that prompt emission from TeV DM may alleviate the discrepancy between the jet model and the TeV data.  Assuming a NFW profile and the presence of DM clumps, they fit the data with a very large value of the annihilation cross-section (typically $3 \times 10^{-24}\ \rm cm^{3}\ s^{-1}$) and a very large boost factor of almost 1000. However, as shown in Fig.~\ref{SSC+DM}, in the presence of a spike we can fit the TeV data for a value of the annihilation cross-section smaller than the thermal value and no additional boost factor is needed.

\begin{table}[b]
\caption{\label{table}Best-fit DM mass and annihilation cross-section, for various characteristic annihilation channels.}
\begin{ruledtabular}
\begin{tabular}{cccc}
Channel &$m_{\mathrm{DM}}\ \rm (TeV)$&$\left\langle \sigma v \right\rangle\ \rm (cm^{3}\ s^{-1})$&$\chi^{2}/\mathrm{d.o.f.}$\\
\colrule
$\bar{b}b$ & $23_{-8}^{+16}$ & $3.9_{-1.4}^{+2.6} \times 10^{-27}$ & 1.34\\
$\tau^+\tau^-$ & $2.1_{-0.5}^{+0.7}$ & $4.4_{-0.9}^{+1.2} \times 10^{-28}$ & 1.30\\
$\bar{q}q$ & $16_{-7}^{+14}$ & $2.7_{-1.2}^{+2.3} \times 10^{-27}$ & 1.46\\
$\bar{t}t$ & $31_{-12}^{+24}$ & $6.0_{-2.3}^{+4.2} \times 10^{-27}$ & 1.33\\
$ZZ$ & $18_{-7}^{+14}$ & $3.9_{-1.5}^{+2.9} \times 10^{-27}$ & 1.29\\
$hh$ & $22_{-8}^{+15}$ & $4.3_{-1.4}^{+2.8} \times 10^{-27}$ & 1.25\\
\end{tabular}
\end{ruledtabular}
\end{table}

Using the Chandra, Fermi-LAT, VERITAS, MAGIC and HESS data (namely 24 data points), our best fit for the jet emission model gives $\chi^{2} \approx 85.5$, that is $\chi^{2}/\mathrm{d.o.f.} \approx 3.56$ for 24 degrees of freedom. Adding a DM contribution (modeled by 2 free parameters, namely the DM mass and annihilation 
cross-section) to this best fit background model considerably improves the quality of the fit. We obtain $\chi^{2} \approx 29.4$ ($\chi^{2}/\mathrm{d.o.f.} \approx 1.34$ for $24-2$ d.o.f) for the $\bar{b}b$ channel (Fig.~\ref{SSC+DM}, left panel) and $\chi^{2} \approx 28.7$ ($\chi^{2}/\mathrm{d.o.f.} \approx 1.30$) for the $\tau^+\tau^-$ channel (Fig.~\ref{SSC+DM}, right panel).  The corresponding best-fit values for the mass and cross-section are given in Table \ref{table}. 

\begin{figure}[t]
\centering
\includegraphics[scale=0.43]{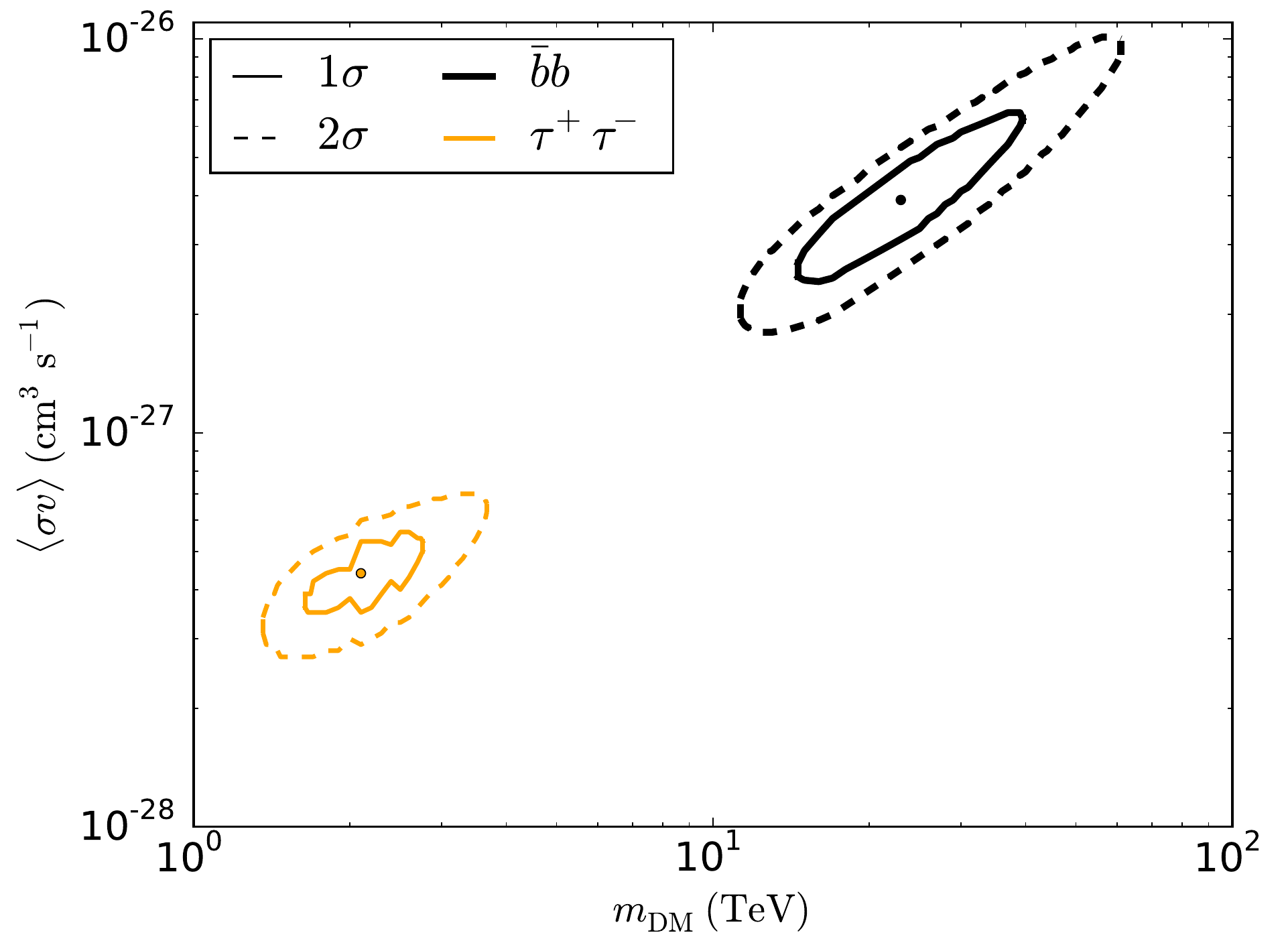}
\caption{\label{contours}Confidence contours at the $1\sigma$ and $2\sigma$ levels, in the plane annihilation cross-section vs DM mass, for the $\bar{b}b$ channel (black thick contours) and the $\tau^+\tau^-$ channel (orange thin contours). The best fit points at the center of the contours correspond to $m_{\mathrm{DM}} = 23\ \rm TeV$, $\left\langle \sigma v \right\rangle = 3.9 \times 10^{-27}\ \rm cm^{3}\ s^{-1}$ for $\bar{b}b$, and $m_{\mathrm{DM}} = 2.1\ \rm TeV$, $\left\langle \sigma v \right\rangle = 4.4 \times 10^{-28}\ \rm cm^{3}\ s^{-1}$ for $\tau^+\tau^-$.}
\end{figure}

The associated $1\sigma$ and $2\sigma$ confidence contour plots for both channels are shown in Fig.~\ref{contours}. For completeness, we also indicate the best-fit values obtained for other annihilation channels. Note that the $\bar{c}c$ and $gg$ channels are degenerate with the $\bar{q}q$ channel, so the same conclusions apply. Similarly $W^+W^-$ and $ZZ$ are also degenerate.

\subsubsection{Dependence on the magnetic field}

The best-fit values displayed in Table \ref{table} are obtained by assuming the same magnetic field intensity as in Sec.~\ref{DM_spike_only} and reflect the fact that very heavy DM candidates give rise to very high energy $\gamma$-rays. However this statement depends on the magnetic field and heavy particles can emit light at much lower energies. For example, we observe that a DM candidate with $m_{\mathrm{DM}} \sim 20\ \rm TeV$  can lead to an excess of X-rays if the magnetic field is relatively weak (typically about $10^{5}-10^{6}\ \mu \rm G$) and be ruled out by the  Chandra data. 

Whether a candidate is ruled out or not, however, also depends on the model for the jet emission. By varying both the jet model and the DM component we can, for example, reconcile a  DM candidate with $m_{\mathrm{DM}} \sim 20\ \rm TeV$   (supposedly ruled out by the Chandra data in the presence of a relatively small magnetic field) with a possible noticeable contribution at TeV energies. Note that for such relatively small values of the magnetic field, ICS and SSC are still negligible.

If the magnetic field is even smaller, typically $\sim 10^{3}\ \mu \rm G$ in the inner region, the synchrotron emission gives a signature at energies of a few eV corresponding to frequencies of about $\sim 10^{15}\ \rm Hz$.  In that case there is no tension with the X-ray data. However, ICS becomes non-negligible for moderate magnetic fields. The SSC emission could also be important but we expect it to be subdominant. For $B \sim 10^{3}\ \mu \rm G$ and $m_{\mathrm{DM}} \sim 20\ \rm TeV$, we expect ICS to give an additional contribution at TeV energies, thus strengthening the case for an explanation of the observed high energy emission in terms of DM.

All these remarks show that the best-fit values obtained by fitting prompt $\gamma$-ray emission give a very good estimate of the contribution of annihilations from a DM spike to the TeV emission, fairly independently of the magnetic field and interstellar radiation field model. Therefore, our  conclusion is that if there is indeed a DM spike in M87, then the subsequent annihilations can account for the TeV $\gamma$-ray emission, with annihilation 
cross-sections 10 times smaller than the thermal value or even smaller depending on the channel.

\section{Conclusion}
\label{conclusion}

We believe that the case for a DM spike at the center of the M87 galaxy is very strong. One therefore expects a significant  
annihilation signal from thermal DM candidates. In this paper, we have confronted the observed SED of M87 with the predicted emission from DM, and set extremely strong upper limits on the annihilation cross-section of DM particles as a function of the DM mass. These limits exclude thermal DM candidates with a velocity-independent (s-wave) cross-section and a mass up to $\mathcal{O}(100) \ \rm TeV$. Our results are independent of the magnetic field distribution and  absorption processes whatever the DM mass; the sole exception is for annihilations into light leptons but our conclusion remains valid for DM masses up to 50 TeV. Also we have shown that in the presence of a DM spike, TeV DM can explain the TeV $\gamma$-ray data for annihilation cross-sections smaller than the canonical value ($\sim 10^{-27}\ \rm cm^{3}\ s^{-1}$).

We expect similar constraints for galaxies containing a supermassive BH with the same mass as in M87.  Should such a spike be found, for example using stellar kinematics, one would exclude a very large chunk of the thermal DM parameter space. This opens up a new path in DM searches, with great potential to elucidate the nature of DM particles. These results provide a strong motivation to look for further evidence for DM spikes in galaxies.

\acknowledgments{We would like to thank Sebastian Wild, Jose A. R. Cembranos, Christoph Weniger, Emmanuel Moulin and Chiara Arina for fruitful discussions and suggestions. This research has been supported at IAP by the ERC Project No.~267117 (DARK) hosted by Universit\'e Pierre et Marie Curie (UPMC) - Paris 6 and at JHU by NSF Grant No.~OIA-1124403. This work has been also supported by UPMC and STFC. Finally, this project has been carried out in the ILP LABEX (ANR-10-LABX-63) and has been supported by French state funds managed by the ANR, within the Investissements d'Avenir programme (ANR-11-IDEX-0004-02). Also, this research, carried out in part during the Dark MALT workshop in Garching, has been supported by the Munich Institute for Astro- and Particle Physics (MIAPP) of the DFG cluster of excellence ``Origin and Structure of the Universe''.}

\section*{Appendix A: DM spike model}

We consider the profile derived by the authors of Ref.~\cite{spikeGS} for a DM spike growing from an initial profile $\propto \rho_{0} \left( r/r_0 \right)^{-\gamma}$:
\begin{equation}
\rho (r) =
\begin{cases}
0 & r < 4 R_{\mathrm{S}} \\
\dfrac{\rho_{\mathrm{sp}}(r) \rho_{\mathrm{sat}}}{\rho_{\mathrm{sp}}(r) + \rho_{\mathrm{sat}}} & 4 R_{\mathrm{S}} \leq r < R_{\mathrm{sp}} \\
\rho_{0} \left( \dfrac{r}{r_0} \right)^{-\gamma} \left( 1 + \dfrac{r}{r_0} \right) ^{-2} & r \geq R_{\mathrm{sp}},
\end{cases}
\end{equation}
where the saturation density determined by DM annihilations reads
\begin{equation}
\rho_{\mathrm{sat}} = \dfrac{m_{\mathrm{DM}}}{\left\langle \sigma v \right\rangle t_{\mathrm{BH}}},
\end{equation}
with $m_{\mathrm{DM}}$ and $\left\langle \sigma v \right\rangle$ respectively the mass and annihilation cross-section of the DM particle, and $t_{\mathrm{BH}}$ the age of the BH. The spike profile reads
\begin{equation}
\rho_{\mathrm{sp}}(r) = \rho_{\mathrm{R}} g_{\gamma}(r) \left( \dfrac{R_{\mathrm{sp}}}{r} \right)^{\gamma_{\mathrm{sp}}},
\end{equation}
where $g_{\gamma}(r) \approx \left( 1 - \frac{4 R_{\mathrm{S}}}{r} \right) ^{3}$, $\rho_{\mathrm{R}} = \rho_{0} \left( \frac{R_{\mathrm{sp}}}{r_0} \right) ^{-\gamma}$, the spike radius is $R_{\mathrm{sp}} = \alpha_{\gamma} r_0 \left( \frac{M_{\mathrm{BH}}}{\rho_{0}r_0^{3}} \right) ^{\frac{1}{3-\gamma}}$ and $\gamma_{\mathrm{sp}} = \frac{9-2\gamma}{4-\gamma}$. We use the values given in Ref.~\citep{Gorchtein_DM_AGN_jet} for the mass of the BH $M_{\mathrm{BH}} = 6.4 \times 10^9\ \mathrm{M_{\odot}}$, the corresponding Schwarzschild radius $R_{\mathrm{S}} = 6 \times 10^{-4}\ \rm pc$, $\alpha_{\gamma} = 0.1$, and $t_{\mathrm{BH}} = 10^{10}\ \rm yr$. We fix $r_0 = 20\ \rm kpc$ for the halo (similarly to the Milky Way), and we must then determine the normalization $\rho_{0}$. 

We choose $\rho_{0}$ in such a way that the profile is compatible with both the total mass of the galaxy and the mass enclosed within the radius of influence of the BH, of order $10^5 R_{\mathrm{S}}$. We thus follow the procedure described in Ref.~\citep{Gorchtein_DM_AGN_jet}: the DM mass within the region that is relevant for the determination of the BH mass, typically within $R_0 = 10^5 R_{\mathrm{S}}$, must be smaller than the uncertainty on the BH mass $\Delta M_{\mathrm{BH}}$. $\rho_0$ is thus obtained by solving the following equation: 
\begin{equation}
\label{normalization}
\int_{4R_{\mathrm{S}}}^{10^5 R_{\mathrm{S}}} \! 4 \pi r^{2} \rho(r) \,\mathrm{d}r = \Delta M_{\mathrm{BH}},
\end{equation}
with $\Delta M_{\mathrm{BH}} = 5 \times 10^8\ \rm M_\odot$. Considering the complex dependence of $\rho$ on $\rho_0$, we use the fact that the mass is dominated by the contribution from $r \gg R_{\mathrm{S}}$, i.e., typically $r > R_{\mathrm{min}} = \mathcal{O}(100 R_{\mathrm{S}})$. In this regime we have $\rho \sim \rho_{\mathrm{sp}}(r)$. We can also factorize the dependence on $\rho_0$ in $\rho_{\mathrm{sp}}$, $\rho_{\mathrm{sp}}(r) = g_{\gamma}(r) \rho_0^{\frac{1}{4-\gamma}} \left( R_{\mathrm{sp}}'/r_0 \right) ^{-\gamma} \left( R_{\mathrm{sp}}'/r \right) ^{\gamma_{\mathrm{sp}}}$, with $R_{\mathrm{sp}}' = \alpha_{\gamma} r_{0} \left( M_{\mathrm{BH}}/r_{0}^{3}\right)^{\frac{1}{3-\gamma}}$, and we finally obtain
\begin{equation}
\rho_0 = \left( \dfrac{\left( 3-\gamma_{\mathrm{sp}}\right) \Delta M_{\mathrm{BH}}}{4 \pi R_{\mathrm{sp}}'^{\gamma_{\mathrm{sp}} - \gamma} r_0^{\gamma} \left( R_0^{3-\gamma_{\mathrm{sp}}} - R_{\mathrm{min}}^{3-\gamma_{\mathrm{sp}}} \right) } \right) ^{4-\gamma}.
\end{equation}
Numerically, we get $\rho_0 \approx 2.5\ \rm GeV\ cm^{-3}$ for $\gamma = 1$. Finally, the total mass within 50 kpc is $\sim 4 \times 10^{12}\ \rm M_{\odot}$, compatible with the value derived from observations, $6 \times 10^{12}\ \rm M_{\odot}$ \citep{
Merritt1993}.

For completeness, we also consider the case of a DM cusp without a spike. In that case the profile is given by:
\begin{equation}
\rho (r) =
\begin{cases}
0 & r < 4 R_{\mathrm{S}} \\
\rho_{\mathrm{sat}} & 4 R_{\mathrm{S}} \leq r < r_{\mathrm{sat}} \\
\rho_{0} \left( \dfrac{r}{r_0} \right)^{-\gamma} \left( 1 + \dfrac{r}{r_0} \right) ^{-2} & r \geq r_{\mathrm{sat}},
\end{cases}
\end{equation}
where $r_{\mathrm{sat}} = r_0 \left( \rho_0/\rho_{\mathrm{sat}} \right) ^{\frac{1}{\gamma}}$, with the same value of $\rho_0$ as in the presence of a spike.

In practice throughout the paper we take $\gamma = 1$, which corresponds to the NFW profile \cite{NFW}. The corresponding spike has a power-law index of $\gamma_{\mathrm{sp}} = 7/3$.

\section*{Appendix B: Synchrotron and prompt emission intensities}

Unless otherwise stated, we used the equipartition magnetic field model, characterized by:
\begin{equation}
\label{B_equi}
B(r) = 
\begin{cases}
B_0 \left( \dfrac{r_{\mathrm{c}}}{r_{\mathrm{acc}}} \right) ^{2} \left( \dfrac{r}{r_{\mathrm{acc}}} \right) ^{-\frac{5}{4}} & r < r_{\mathrm{acc}} \\
B_0 \left( \dfrac{r}{r_{\mathrm{c}}} \right) & r_{\mathrm{acc}} \leq r < r_{\mathrm{c}} \\
B_0 & r \geq r_{\mathrm{c}}.
\end{cases}
\end{equation}
We take $B_0 = 10\ \rm \mu G$ for the large scale value of the magnetic field outside the inner cocoon of radius $r_{\mathrm{c}} \sim 10\ \rm kpc$ seen for instance by LOFAR \cite{M87_LOFAR}. The authors of Ref.~\cite{Regis&Ullio} estimate the radius of the accretion region as $r_{\mathrm{acc}} = 2 G M_{\mathrm{BH}}/v_{\mathrm{flow}}^{2}$, where $v_{\mathrm{flow}} \sim 500 - 700\ \rm km\ s^{-1}$ is the velocity of the Galactic wind at the center of the Milky Way. Here we assume similar characteristics for the wind at the center of M87, so we just rescale the BH mass. For the Milky Way, the size of the accretion region was $\sim 0.04\ \rm pc$. Now, considering that the black hole in M87 has a mass approximately $1.5 \times 10^{3}$ times larger than the one at the center of the Milky Way, Sgr A*, we estimate $r_{\mathrm{acc}} \sim 60\ \rm pc$. The resulting equipartition magnetic field can reach very large values at the center, typically up to $10^{10}-10^{11}\ \rm \mu G$ in the very inner region.

To compute the synchrotron intensity, we first need the electron (and positron) spectrum, given by (see, e.g., \cite{Regis&Ullio})
\begin{equation}
\psi_{\mathrm{e},i}(r,E) = \dfrac{1}{b(r,E)} \dfrac{\left\langle \sigma v \right\rangle_{i}}{\eta} \left( \dfrac{\rho(r)}{m_{\mathrm{DM}}} \right) ^{2} \int_{E}^{m_{\mathrm{DM}}} \! \dfrac{\mathrm{d}N_{\mathrm{e},i}}{\mathrm{d}E_{\mathrm{S}}} \left( E_{\mathrm{S}} \right) \, \mathrm{d}E_{\mathrm{S}},
\end{equation}
where we use $\eta = 2$---which corresponds to the assumption that the DM particle is a Majorana fermion---and $\mathrm{d}N_{\mathrm{e},i}/\mathrm{d}E_{\mathrm{S}}$ is the electron or positron injection spectrum that we take from Ref.~\cite{Cirelli_cookbook} for each channel denoted by $i$. We use injection spectra that include electroweak corrections that become very important for large masses. $b(r,E)$ is the total energy loss rate. Since we consider large magnetic fields in the inner region, essentially all the energy of electrons and positrons is lost in the form of synchrotron radiation and the contributions from ICS and synchrotron self-Compton turn out to be negligible \cite{Aloisio2004}, so that the loss term reads (see, e.g.~\cite{Longair2011})
\begin{equation}
b(r,E) = b_{\mathrm{syn}}(r,E) = \dfrac{4}{3} \sigma_{\mathrm{T}} c \dfrac{B(r)^{2}}{2 \mu_{0}} \gamma_{\mathrm{L}} ^{2},
\end{equation}
where $ \sigma_{\mathrm{T}} $ is the Thomson cross-section, $ B(r) $ the intensity of the magnetic field, $c$ the speed of light, $\gamma_{\mathrm{L}} = E/(m_{\mathrm{e}} c^{2})$ the Lorentz factor of the electrons, $m_{\mathrm{e}}$ the electron mass and $ \mu_{0} $ the vacuum permeability. 

Then the synchrotron emissivity reads (see e.g.~Ref.~\cite{Fornengo2012_syn})
\begin{equation}
j_{\nu,i}(r) = 2 \int_{m_{\mathrm{e}}}^{m_{\mathrm{DM}}} \! P_{\nu}(r,E) \psi _{\mathrm{e},i}(r,E) \, \mathrm{d}E, 
\end{equation}
where the factor 2 refers to the fact that both an electron and a positron are produced in one DM annihilation, and the synchrotron emission spectrum is given by (see, e.g., Ref.~\cite{Longair2011})
\begin{equation}
P_{\nu}(r,E) = \dfrac{1}{4 \pi \epsilon_{0}} \dfrac{\sqrt{3}e^{3}B(r)}{m_{\mathrm{e}}c} G_{\rm{i}} \left( \dfrac{\nu}{\nu_{\mathrm{c}}(r,E)} \right),
\end{equation}
where $ e $ the elementary charge, $ \epsilon_{0} $ the vacuum permittivity, and the critical frequency is given by
\begin{equation}
\nu_{\mathrm{c}}(r,E) = \dfrac{3eE^{2}B(r)}{4 \pi m_{\mathrm{e}}^{3}c^{4}}.
\end{equation}
$G_{\rm{i}}$ is the isotropic synchrotron spectrum, obtained by averaging the synchrotron spectrum over an isotropic distribution of pitch angles \cite{Longair2011}:
\begin{equation}
G_{\rm{i}}(x) = \dfrac{1}{2} \int_{0}^{\pi} \! G\left( \dfrac{x}{\sin \alpha} \right) \sin^{2}\alpha \, \mathrm{d}\alpha,
\end{equation}
with $G(t) = t \int_t^\infty \! K_{5/3}(u) \, \mathrm{d}u$, where $K_{5/3}$ is the modified Bessel function of order 5/3. To simplify the numerical treatment of the angle average, one may use, for instance, the parametrization described in Ref.~\cite{syn_parametrization}.

From there, the specific intensity for a given angle $\theta$ from the center is given by the integral of the emissivity over the line of sight (l.o.s.\!):
\begin{equation}
\label{I_nu_syn_appendix}
I_{\nu,i}^{\mathrm{syn}}(\theta) = \int_\mathrm{l.o.s.} \! \dfrac{j_{\nu,i}(r(s,\theta))}{4 \pi} \, \mathrm{d}s,
\end{equation}
$s$ being the radial coordinate along the line of sight and $r(s,\theta) = \sqrt{d^{2} + s^{2} - 2ds\cos\theta} \approx \sqrt{(d - s)^{2} + ds\theta^{2}}$. The approximation of small angles is justified since the characteristic radius $R_{\mathrm{M87}}$ of M87 (typically 50 kpc) is much smaller than the distance of M87, $d = 16\ \rm Mpc$. Also in practice we perform the integral over the l.o.s.~between $d - R_{\mathrm{M87}}$ and $d + R_{\mathrm{M87}}$, considering the concentrated nature of the DM profile.

\begin{figure}[t]
\centering
\includegraphics[scale=0.45]{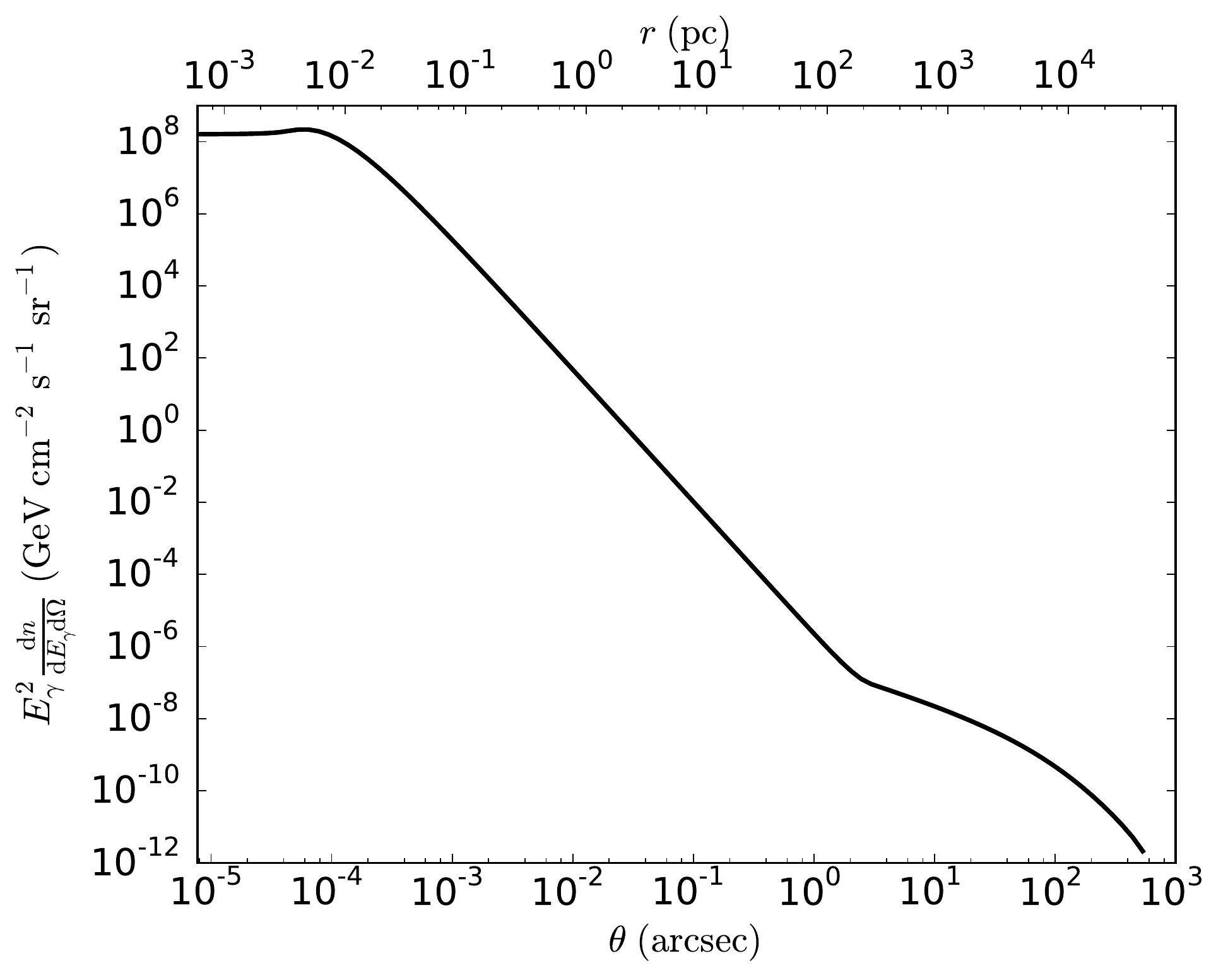}
\caption{\label{I_nu_1TeV}$\gamma$-ray specific intensity at 1 TeV as a function of the angle from the center, for a DM spike with $\gamma_{\mathrm{sp}} = 7/3$, for the $\bar{b}b$ channel, $m_{\mathrm{DM}} = 23$ TeV and $\left\langle \sigma v \right\rangle = 3.9 \times 10^{-27}\ \rm cm^{3}\ s^{-1}$.}
\end{figure}

The specific intensity for prompt $\gamma$-rays is simply given by the integral over the line of sight of the DM density squared (see, e.g., Ref.~\cite{Bringmann_review} and references therein),
\begin{equation}
\label{I_nu_prompt_appendix}
\nu I_{\nu,i}^{\mathrm{prompt}}(\theta) \equiv E_{\gamma}^{2}\dfrac{\mathrm{d}n_{i}}{\mathrm{d}E_{\gamma}\mathrm{d}\Omega} = \dfrac{E_{\gamma}^{2}}{4\pi \eta} \dfrac{\left\langle \sigma v \right\rangle_{i}}{m_{\mathrm{DM}}^{2}} \dfrac{\mathrm{d}N_{\gamma,i}}{\mathrm{d}E_{\gamma}} \int_{\mathrm{l.o.s.}} \! \rho^{2}\left( r(s,\theta)\right)  \, \mathrm{d}s,
\end{equation}
where $\mathrm{d}N_{\gamma,i}/\mathrm{d}E_{\gamma}$ is the prompt gamma ray spectrum taken from Ref.~\cite{Cirelli_cookbook} and $E_{\gamma} = h \nu$ with $h$ the Planck constant. The specific intensity at 1 TeV as a function of the angle from the center, for a DM spike with $\gamma_{\mathrm{sp}} = 7/3$, for annihilations proceeding to $\bar{b}b$, with $m_{\mathrm{DM}} = 23$ TeV and $\left\langle \sigma v \right\rangle = 3.9 \times 10^{-27}\ \rm cm^{3}\ s^{-1}$, is shown in Fig.~\ref{I_nu_1TeV}. The integral over the l.o.s.~is approximately constant at the center, due to the vanishing density below $4 R_{\mathrm{S}}$, hence the plateau below $\sim 100\ \mu \rm as$. Above $\sim 1\ \rm arcsec$, the change in slope is related to the outer part of the DM profile, assumed to follow the NFW distribution.

\section*{Appendix C: Synchrotron self-Compton model for the jet}

Here, we summarize the leptonic SSC model for the spectral energy distribution of M87 described in Ref.~\cite{SSC_model} and used by the Fermi Collaboration in Ref.~\cite{Fermi_M87}. All the primed quantities are defined in the rest frame of the plasma blob. Considering that the redshift of M87 is 0.00428,\footnote{\url{http://messier.seds.org/xtra/supp/m_NED.html}} we neglect redshift effects in our discussion. Following the notations of Ref.~\cite{SSC_model}, the observed synchrotron flux $\nu F_{\nu}$ is denoted $f_{\epsilon}$, where $\epsilon = h \nu/(m_{\mathrm{e}} c^{2})$ is the dimensionless energy of the emitted synchrotron in the observer's frame. Similarly to the synchrotron flux obtained in Appendix B, the observed flux, rewritten in terms of dimensionless energies, and taking into account the Doppler boost, reads:
\begin{equation}
f_{\epsilon}^{\mathrm{syn}} = \dfrac{\sqrt{3} \delta_{\mathrm{D}}^{4} \epsilon' e^{3} B c}{4 \pi h d^{2}} \int_1^{\infty} \! N_{\mathrm{e}}'(\gamma') G_{\mathrm{i}}(x(\epsilon',\gamma')),
\end{equation}
where $\delta_{\mathrm{D}}$ is the Doppler factor, $\epsilon' = \epsilon/\delta_{\mathrm{D}}$, and $d = 16\ \rm Mpc$ is the distance of M87. $N_{\mathrm{e}}' = n_{\mathrm{e}}' V'_{\mathrm{b}}$ is the electron distribution in the rest frame of the blob, with $n_{\mathrm{e}}'$ the electron number density, and $V'_{\mathrm{b}}$ the volume of the blob. $x = \nu'/\nu'_{\mathrm{c}}$ is rewritten in terms of the dimensionless quantities:
\begin{equation}
x(\epsilon',\gamma') = \dfrac{4 \pi \epsilon' m_{\mathrm{e}}^{2} c^{2}}{3 e B h \gamma'^{2}}.
\end{equation}
Note that the intensity of the magnetic field $B$ is not primed but is also defined in the rest frame of the blob. For $G_{\mathrm{i}}$, we take the parametrization of Ref.~\cite{syn_parametrization}, also used in Ref.~\cite{SSC_model}. For the electron distribution, we consider as in Ref.~\cite{Fermi_M87} a broken power law:
\begin{equation}
N_{\mathrm{e}}'(\gamma') = K
\begin{cases}
\gamma'^{-p_{1}} & 1 \leq \gamma' \leq \gamma'_{1} \\
\gamma_{1}'^{p_2-p_1} \gamma'^{-p_{2}} & \gamma'_{1} < \gamma' \leq \gamma'_{2} ,
\end{cases}
\end{equation}
where $\gamma'_{1} = 4 \times 10^{3} $ is the Lorentz factor at the break and $\gamma'_{2} = 10^{7}$ is the maximum Lorentz factor of the electrons.

From there, the observed SSC flux is given by \cite{SSC_model}
\begin{equation}
\label{SSC_flux}
f_{\epsilon_{\mathrm{s}}}^{\mathrm{SSC}} = \dfrac{9 \sigma_{\mathrm{T}} \epsilon_{\mathrm{s}}'^{2}}{16 \pi \delta_{\mathrm{D}}^{2} c^{2} t_{\mathrm{v,min}}^{2}} \int_0^{\infty} \! \dfrac{f_{\epsilon}^{\mathrm{syn}}}{\epsilon'^{3}} \int_{\gamma'_{\mathrm{min}}}^{\gamma'_{\mathrm{max}}} \! \dfrac{N_{\mathrm{e}}'(\gamma')}{\gamma'^{2}} F_{\mathrm{C}}(q,\Gamma) \, \mathrm{d}\gamma' \, \mathrm{d}\epsilon',
\end{equation}
where $t_{\mathrm{v,min}} = R'_{\mathrm{b}}/(\delta_{\mathrm{D}} c)$ is the variability time scale of the source, $R'_{\mathrm{b}}$ being the (comoving) radius of the blob. $\epsilon_{\mathrm{s}}'$ is the dimensionless energy of the scattered photon. The ICS process is encoded in $F_{\mathrm{C}}(q,\Gamma)$ which reads
\begin{equation}
F_{\mathrm{C}}(q,\Gamma) = 2 q \ln q + (1 + 2 q) (1 - q) + \dfrac{(\Gamma q)^{2}}{2(1 + \Gamma q)} (1 - q)
\end{equation}
if $1/(4\gamma'^{2}) \leq q \leq 1$ and $F_{\mathrm{C}}(q,\Gamma) = 0$ otherwise. $q$ and $\Gamma$ are given by:
\begin{equation}
q = \dfrac{\epsilon_{\mathrm{s}}'/\gamma'}{\Gamma (1 - \epsilon_{\mathrm{s}}'/\gamma')},\ \ \ \Gamma = 4 \epsilon' \gamma'.
\end{equation}
The kinematically allowed range of values for $q$ translates into the integration bounds in Eq.~\eqref{SSC_flux}:
\begin{equation}
\gamma_{\mathrm{min}}' = \dfrac{1}{2} \epsilon_{\mathrm{s}}' \left( 1 + \sqrt{1 + \dfrac{1}{\epsilon' \epsilon_{\mathrm{s}}'}} \right),
\end{equation}
\begin{equation}
\gamma_{\mathrm{max}}' = 
\begin{cases}
\dfrac{\epsilon' \epsilon_{\mathrm{s}}'}{\epsilon' - \epsilon_{\mathrm{s}}'} & \epsilon' > \epsilon_{\mathrm{s}}' \\
\gamma'_{2} & \epsilon' \leq \epsilon_{\mathrm{s}}'.
\end{cases}
\end{equation}

\newpage

\bibliographystyle{h-physrev} 
\bibliography{/Users/thomaslacroix/Documents/thesis/papers/biblio/biblio}

\end{document}